\documentclass{aa}

\usepackage{graphicx}
\usepackage{color}
\usepackage{txfonts}
\begin{document}

   \title{Cosmological-scale coherent orientations \\
of quasar optical polarization vectors in the \textit{Planck} era}

   \subtitle{Investigating the Galactic dust contamination scenario}

\titlerunning{Quasar optical polarization alignments in the \textit{Planck} era}

   \author{V. Pelgrims
          \inst{1}
          }

   \institute{Univ. Grenoble Alpes, CNRS, Grenoble INP, LPSC-IN2P3, 38000 Grenoble, France\\
              \email{pelgrims@lpsc.in2p3.fr}
             }

   \date{Received May XX, 2017; accepted mmmmm XX, YYYY}

  \abstract{
   Gigaparsec scale alignments of the quasar optical polarization
   vectors have been proven to be robust against a scenario of
   contamination by the Galactic interstellar medium (ISM). This
   claim has been established by means of optical polarization
   measurements of the starlight surrounding the lines of sight of
   the 355 quasars for which reliable optical polarization
   measurements are available.
   In this paper, we take advantage of the full-sky and high quality
   polarization data released by the \textit{Planck} satellite to
   provide an independent, complementary, and up-to-date
   estimation of the contamination level of the quasar optical
   polarization data by the Galactic dust.
   Our analysis reveals signatures of Galactic dust contamination
   at the two sigma level for about 30 per cent of the quasar optical
   polarization data sample. The remaining 70 per cent of the lines
   of sight do not show Galactic dust contamination above the two
   sigma level, suggesting low to negligible contamination of the
   quasar optical polarization signal.
   We further found arguments suggesting that Galactic thermal dust
   cannot fully account for the reported quasar optical polarization
   alignments.
   Based on the measurements of the ratio of the polarized intensity of
   the dust in the submillimeter to the degree of linear polarization of
   the quasar in the optical, we provide a new and independent quality
   criteria to apply to the quasar optical polarization sample.
   We argue that, unless correction is applied, such a criterion should
   be imposed on the data for future investigations in the framework of
   the cosmological-scale correlations of quasar optical polarization
   vector orientations that still could compete with the isotropic principle
   of the cosmological paradigm.
   }

   \keywords{polarization --
                quasar: general --
                submillimeter: ISM --
                ISM: dust, magnetic field --
                large-scale structures of Universe
               }

   \maketitle
%

\section{Introduction}
\citeauthor{Hut2005} (\citeyear{Hut1998,Hut2001,Hut2005}) have
reported alignments of quasar optical linear polarization vectors
extending over cosmological-scale regions of the Universe (see
also \citealt{Jain2004,Pel2014,Hut2014,Pel2015,Pel2016,myThesis2016}
for a recent review).
The very large scales at which these coherent orientations take place
suggest a cosmological origin or at least a correlation between objects
or fields over cosmological scales, up to $\sim$ 1 Gpc at redshift
$z \simeq 1 - 2$.
These observations were recognized as reflecting the need to
add new ingredients to the well-accepted cosmological concordance
model.
Since their observational discovery, a wide variety of scenarios have
been proposed to explain these alignments.
A nonexhaustive list of these includes fundamental constant variations,
cosmic strings, cosmic birefringence, cosmological magnetic fields,
new dark matter particle candidates, anisotropic or rotational cosmological
models, and bad sampling or contamination of the data set
(see, e.g., \citealt{Cha2012,Pol2010,diS2015,Urb2010,Tiw2015,
Das2005,Pay2008,Hut2011,Aga2011,Hut2005,Cia2012,Kuv2014,
Jos2007}).
It is worth mentioning that none of these scenarios have been able yet
to reproduce fully the observational characteristics of these alignments
(e.g., \citealt{Hut2010,diS2015}).

Because the polarization vectors of objects located in the same
region of the sky but at different redshift do not show the same
alignment direction, it has been argued that possible instrumental
bias and contamination by the interstellar medium (ISM) are unlikely
to be responsible for the observed polarization angle correlations
(\citeauthor{Hut2005} \citeyear{Hut1998,Hut2001,Hut2005}, and \citealt{Pay2010}).
The eventuality of instrumental bias has been discussed by
\citeauthor{Hut2005} (\citeyear{Hut1998,Hut2005}) and \citet{Slu2005}.
Briefly, their conclusion is that instrumental bias is very unlikely
because the quasar optical polarization sample contains data from
different observational campaigns, measured by different authors, using
different techniques and/or with different instruments.
Additionally, the global significance of the patterns of polarization
alignments has been assessed using dedicated statistical techniques,
specifically based on the reshuffling of the polarization vectors on the
source positions that would have taken a global bias into account when
computing the prior of randomness (see \citealt{Jain2004,Hut2005}).

As discussed further in Sect.~\ref{sec:Sect_3},
the combination of magnetic fields and dust grains in our Galaxy is
well known to polarize optical light from distant stars and to be at the
origin of alignment patterns of star polarization vectors (e.g.,
\citealt{Mat1970,Axo1976,PlanckXXI2015,PlanckXII2018}).
While the quasar light is certainly contaminated to some degree by
the traversed ISM, \citeauthor{Hut2005} (\citeyear{Hut1998,Hut2005})
made sure it cannot be at the origin of the large-scale quasar
polarization vector alignments.
First, they used \citet{Bur1982} extinction maps to evaluate
the contamination of the quasar polarization data by the ISM. As a
consequence they imposed severe selection criteria on the quasar
polarization data to eliminate at best potentially strong contaminated
data points.
Second, they considered the polarized Galactic star catalogs of
\citet{Axo1976} and \citet{Hei2000} to compare the quasar polarization
vector orientations with those of the nearest polarized Galactic stars.
The detection of the large-scale alignments of the quasar polarization
vectors were found to be robust against the scenario of a Galactic
dust contamination. A similar conclusion was also found in
\cite{Pay2010} where a basic modeling of the ISM contamination was
discussed.

In this paper, our aim is to provide independently and with
up-to-date data sets a new estimate of the degree of contamination
of the quasar optical polarization sample by the Galactic
dust and thus to test further the possibility following which Galactic dust
could be at the origin of the reported alignments.
For this purpose, we make use of the
diffuse thermal dust polarized emission measured by the \textit{Planck}
satellite\footnote{http://www.esa.int.Planck} at 353 GHz
\citep{PlanckXIX2015}.
Compared to previous studies that rely on starlight polarization,
the use of the \textit{Planck} polarized dust data has the considerable
advantage of probing the whole line of sight through the Milky Way. Some
differences between the results obtained using starlight polarization and
thermal dust are thus to be expected.
Another advantage of considering the maps of the dust polarized
emission is from the angular resolution. We may indeed probe rapid
spatial variations (on sky) of dust column density and/or of the grain alignment
direction that would have been otherwise missed using sparse starlight
polarization data.
This analysis should yield independent quality criteria for the quasar
data and possibly allows, in the future, for the refinement of our
understanding of these striking alignments of the quasar polarization
vectors.

\medskip

The paper is structured as follow.
In Sect.~2, we present the data
sets that we use to evaluate the ISM contamination of the quasar
optical polarization data.
Sections~3 and~4 contain the core of our analysis and we present and discuss
thereby the main results with caveats. We conclude in Sect.~5 and
give perspective for future work.
Appendix~A contains details about our treatment of the dust polarization
maps. The robustness of our analysis regarding the choice of the
adopted smoothing length of the polarization maps is discussed in
Appendices~B and~C. In Appendix~D, we reproduce the main results
presented in the rest of the paper but obtained by means of another
polarization ratio discussed by \citet{Mar2007} and used in
\cite{PlanckXXI2015}.


\section{Data sets}

\subsection{Quasar optical polarization catalog}
The current sample of quasar optical polarization measurements
from which the very large-scale alignments of the polarization vectors
are studied is made of 355 sources.
This sample contains reliable polarization measurements for quasars
located in both Galactic hemispheres: 195 quasars are located in the
northern hemisphere and 160 in the southern hemisphere. The quasars
that make the sample have been selected to avoid at best the
ISM Galactic contamination according to the following requirements:
$| b | \geq 30^\circ$,
$p_{\rm{V}} \geq 0.6 \%$ and $\sigma_{\rm{\psi_V}} \leq 14^\circ$,
which corresponds to about  $p_{\rm{V}} \geq 2\, \sigma_{\rm{p_{V}}}$;
$b$ is the Galactic latitude of the source; $p_{\rm{V}}$
the degree of linear polarization; $\sigma_{\rm{p_{V}}}$
its error; and $\sigma_{\rm{\psi_V}}$ the error on the polarization position
angle. We denote the quasar optical polarization position angle
$\psi_{\rm{V}}$, which is defined in the IAU convention (east of north)
and is expressed in degree. It takes values in the range 0 -- 180$^\circ$.
The subscript ${\rm{V}}$ is used throughout for quasar optical polarization
quantities. This choice also indicates that the quasar data are from the
optical domain, mostly in the V band \citep{Slu2005}.

The redshift of the quasars of the final sample varies from 0.06 to 3.95.
Even though all Galactic longitudes are covered, the sky coverage
of the sample is unfortunately not homogeneous
(see \citealt{Hut2005,myThesis2016} for details).
Up to now and to the best of our knowledge it is the only publicly
available\footnote{http://cdsarc.u-strasbg.fr/viz-bin/qcat?J/A+A/441/915}
large and reliable sample of linear polarization data of quasars at optical
wavelengths. \citet{Hut2005} provide a detailed
description of this sample, which consists in a compilation of ``new''
observations obtained at that time (\citealt{Hut1998,Hut2001,Hut2005,Slu2005})
and of ancillary data from the literature.

To proceed to our analysis, we first convert the quasar sky coordinates
from B1950 to the J2000. Then, we convert them into Galactic coordinates
$(l,\,b)$.
The quasar polarization position angles are transformed accordingly using,
for example, Eq. 16 of \citet{Hut1998}.

\subsection{\textit{Planck} 353 GHz polarized sky and dust data}
The diffuse thermal dust polarized emission is the dominant Galactic
foreground present in measurements of the polarization of the cosmic
microwave background (CMB) emission at frequencies above 100 GHz
(e.g., \citealt{PlanckXXII2015}). The \textit{Planck} satellite recorded this
emission and it provided unprecedented full-sky maps of this
Galactic emission in intensity $I_{\rm{S}}$ and polarization, measuring
the $Q_{\rm{S}}$ and $U_{\rm{S}}$ Stokes parameters.
We use the subscript ${\rm{S}}$ to refer to the polarization quantities
in the submillimeter.

We use the \textit{Planck} second release of the single-frequency polarization
maps at 353 GHz that are available on the Planck Legacy
Archive\footnote{http://pla.esac.esa.int/pla/\#maps}.
\citealt{PlanckXIX2015} and \citealt{PlanckXXI2015} give details
and discussions regarding these data.
The \textit{Planck} DR2 HFI 353 GHz maps have a native
resolution\footnote{https://wiki.cosmos.esa.int/planckpla2015/index.php/HFI\_
\\ performance\_summary}
of about 4.94' \citep{PlanckXIX2015} and a
HEALPix\footnote{See http://healpix.sourceforge.net}
grid tessellation corresponding to $N_{\rm{side}} = 2048$ \citep{Gor2005}.
At the instrument resolution, the 353-GHz polarization maps are noise
dominated \citep{PlanckXIX2015}.
This is particularly true at high Galactic latitudes ($| b | \geq 30^\circ$)
where the ISM is more diffuse and has a low density. In order to increase
the signal-to-noise ratio (S/N) of the \textit{Planck} HFI measurements of the
diffuse ISM we smooth the Stokes parameter maps.
However, smoothing the \textit{Planck} data accentuates the difference
between the dust beam relative to the quasar pencil beam.
From a physical point of view, the larger the smoothing value we adopt,
the larger the angular scales of the ISM variations we consider as we
dilute any small-scale spatial structures which, however, might be
important in our case. Therefore, for our comparative analysis to
be meaningful, there is a compromise to be reached between
maintaining high resolution data and dealing with reliable dust
polarization quantities. This is especially true in our case given that
smoothing of the maps could lead to lower polarized intensities due
to beam depolarization and, consequently, biases in our results.

As shown in Appendix~A, we found that it was convenient to smooth the polarization
maps with a Gaussian kernel of full width at half maximum (FWHM) of 15'
; owing to the intrinsic resolution of the map, the effective beam is then about 15.8'.
In order to check the robustness of our results regarding the adopted
smoothing kernel value, we also conducted the analyses presented
in Sect.~3 and~4 in parallel using a smoothing kernel with FWHM of 5', 10', and 20'.
Globally, the results with the different kernel FWHM values are very
consistent as shown in the appendices.

\medskip

At 353 GHz, thermal dust is, by large, the principal source of the polarized
signal. The dispersion arising from CMB polarization anisotropies is much
lower than the instrumental noise for $Q_{\rm{S}}$ and $U_{\rm{S}}$
\citep{PlanckVI2014}. Its impact on our analysis is thus expected
to be negligible.
The cosmic infrared background (CIB) is assumed to be unpolarized
\citep{PlanckXXX2014}.
At this frequency, only small contributions are expected in the intensity
map from the CMB,  CIB, and zodiacal light (e.g.,
\citealt{PlanckXXI2015}). As we used only the intensity map to reinforce other
results, we found that it was unnecessary to correct the map from these contributions.

\medskip

For each quasar we identified the HEALPix pixel corresponding to its
position on the sky and we estimated the thermal dust emission quantities
from the smoothed maps. We note that quasars of our sample are point-like
sources having angular size below the arcsecond level \citep{Slu2005}.
Their contributions, or the contributions of their host Galaxy, to the 353 GHz
signal collected by \textit{Planck} are considered as negligible because of the
instrument resolution at 353 GHz, which is the
scale at which diffuse Galactic components dominate largely the signal.
Their contributions are indeed diluted by the \textit{Planck} beam, a
situation that is enhanced by the additional smoothing of the data that we
performed.

From the $Q_{\rm{S}}$ and $U_{\rm{S}}$ of the smoothed
maps, we derived the dust (linear) polarized intensity $P_{\rm{S}}$,
computed as $P_{\rm{S}}^2 = Q_{\rm{S}}^2 + U_{\rm{S}}^2$, and $\psi_S$
the dust polarization position angle that we define as
\begin{equation}
\psi_S = \frac{1}{2}\, \arctan(- U_{\rm{S}},\,Q_{\rm{S}})\,;
\end{equation}
the minus sign accounts for the conversion of the position angles from
the HEALPix (or COSMO) convention to the IAU convention (east of north).
We forced the polarization position angles to be in the range 0 -- 180$^\circ$.
The evaluation of the errors on $P_{\rm{S}}$ and $\psi_S$ relies on
simulations as explained in Appendix~A.

\medskip

For our analysis, we also made use of the full-sky dust extinction map
derived in \cite{PlanckXI2014}, which provides discussions and details. This quantity,
directly related to the optical depth of the ISM toward the quasars, helps
to evaluate the level of contamination of the quasar optical light by the
Galactic dust (see \citealt{PlanckXXI2015} and our Sect. 3).
For consistency, we also smoothed the $E(B-V)$ map with the same
Gaussian kernel FWHM values as for the polarization maps.

\section{Dust contamination scenario: Analysis and results}
\label{sec:Sect_3}
In the optical, the principal source of polarization contamination from
the Galaxy is due to thermal dust (e.g.,
\citealt{PlanckXIX2015,PlanckXX2015,PlanckXXI2015}).
If the magnetic field is coherent in a given region of the sky, dust grains
align with the lines of the field in that region (\citealt{Mar2007} and
references therein for viable physical models producing the alignment).
The polarization state of an incident light beam passing through this
region is then modified. The electric component perpendicular to the
magnetic field is anisotropically dimmed by diffusion on the dust grains.
An unpolarized incident light beam therefore exits the dusty region with
a net polarization parallel to the magnetic field.
The diffuse Galactic dust also emits thermally polarized light in the
submillimeter spectral band. The polarization of this light is to be
preferentially perpendicular to the lines of the magnetic field in which
the dust grains are embedded (see, e.g., \citealt{Mar2007} for a
discussion).

As a consequence, for a given line of sight, polarization vectors in the
submillimeter and in the visible are expected to be orthogonal to one
another if the optical polarization is only due to dust.
If the background quasar has an intrinsic polarization then the dust
contamination can be simplistically modeled as the vectorial addition
of an ISM polarization (perpendicular to the dust polarized emission)
to the intrinsic quasar polarization. An offset toward the perpendicularity
can, in principle, be detected statistically.
That excess is expected to be better detected the larger the dust-induced
optical polarization compared to the quasar intrinsic degree of optical
polarization.

Below we introduce the quantities that link dust and quasar polarization
data and with which we intend to estimate the eventual contamination of the
quasar optical polarization by the intervening Galactic dust.

\subsection{Dust-to-quasar polarization correlation quantities}
\label{subsec:3.1}
\begin{figure}
        \centering
        \includegraphics[width=\hsize]{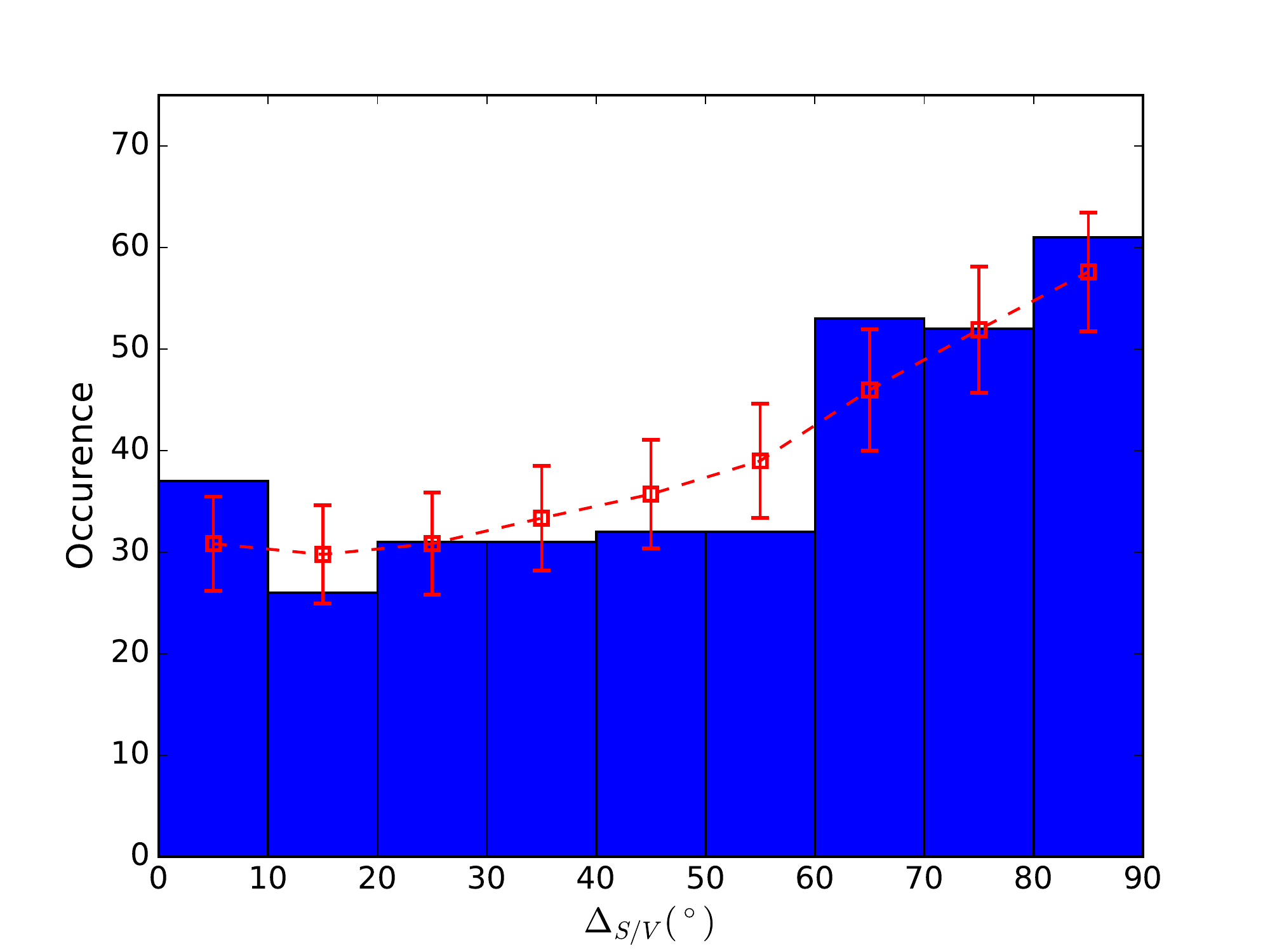}
        \caption{
        Histogram of the 355 relative angles $\Delta_{\rm{S/V}}$ computed as
        in Eq.~\ref{eq:DeltaPsi}. The observed distribution is significantly not
        uniform and shows an excess toward 90 degrees, which corroborates
        a contamination scenario.
        The central values and error bars given in red account for observational
        uncertainties. The errors are estimated through $10^6$ simulations in which
        we vary the polarization position angles of the quasars and dust
        around their values according to the observational uncertainties
        (see text for details). The central values and error bars correspond
        to the mean and dispersion in each bin obtained by computing the histogram
        for each simulation.
        The average correlation appears weaker because of the blurring of the correlation
        that the randomization of the angles within their uncertainties produces.
        }
        \label{fig:histDelta}
\end{figure}

As discussed above, optical polarization of a background source that
would be only due to depolarization by dust is predicted to be
perpendicular to the submillimeter polarization emitted by the dust
grains.
The relative angle between the measured polarization vectors in the
two wavebands can thus be considered as an indicator for dust
contamination of the quasar optical polarization data.
The relative angle between the two polarization vectors is computed
as
\begin{equation}
\Delta_{\rm{S/V}} = 90 - | 90 - |\psi_{\rm{S}} - \psi_{\rm{V}}| |
\label{eq:DeltaPsi}
,\end{equation}
where the angles are expressed in degree. The consecutive absolute
values take into account the axial nature of the polarization vectors. The value
$\Delta_{\rm{S/V}}$ is defined in the range 0 -- 90$^\circ$.
For a sample of lines of sight, any deviation from uniformity of the
$\Delta_{\rm{S/V}}$ distribution toward 90$^\circ$ could indicate a
possible contamination of the sample. Therefore, we consider the
$\Delta_{\rm{S/V}}$ as our first dust-to-quasar contamination gauge.
The $\Delta_{\rm{S/V}}$ distribution of the 355 comparison
measurements is shown in Fig.~\ref{fig:histDelta}.
A clear departure from uniformity is seen. We quantify and discuss this in
the next subsection.
We note that a departure from uniformity would not provide a sufficient
argument to conclude for a contamination. Coincidental position angle
correlations need to be taken into account.
Such consideration is explored in Sect.~3.5.
We also note that the opposite is equally true; a uniform distribution of the
$\Delta_{\rm{S/V}}$ is not a sufficient argument to claim for no contamination.
For example, large experimental uncertainties can
uniformize the angle distribution or the dust-induced polarization could
be not sufficient to override completely the intrinsic optical polarization
of the background source (see also the discussion in Sect.~4.2.3). As a
result, the criterion based on the
$\Delta_{\rm{S/V}}$ distribution can only be used to diagnose a
significant contamination of the optical light by the intervening dust.

\medskip

We complement our study using two polarization ratios discussed in
\cite{Mar2007} and that were recently used in \cite{PlanckXXI2015} to
derive ISM physical properties by comparing polarized light from stars in
the optical domain (V band) and thermal dust emission from the same
\textit{Planck} data that we use in this work. 

\medskip

The first polarization ratio we consider is defined as
\begin{equation}
R_{\rm{P/p}} = P_{\rm{S}} / p_{\rm{V}}
\end{equation}
and has units of polarized intensity (here K${}_{\rm{CMB}}$).
For Galactic star-related studies, this ratio is used to characterize the
efficiency of the dust grains at producing polarized submillimeter
emission compared to their ability at polarizing starlight in the visible.
In this case, the fundamental difference with starlight based studies is that the
quasars of our sample are thought to have an intrinsic degree of linear
polarization while the starlight are assumed to be unpolarized.
The $R_{\rm{P/p}}$ values are therefore expected to be lower in
our study than in Planck Collaboration XXI (2015), for example.
Also a consequence of this is that the polarization ratio is defined
even when $P_{\rm{S}}$ is zero.

Large values of the $R_{\rm{P/p}}$ require at the same time a large
polarized dust emission and a low degree of linear polarization of the
quasars. The first condition implies dusty regions with a significant
alignments of nonspherical dust grains which, in turn, implies an
efficient polarizability of incident optical light.
The second condition is fulfilled for quasars that are more prone to be
affected by dust contamination,
i.e., those with the smallest intrinsic degrees of linear polarization.

\medskip

The other polarization ratio that we consider is defined as
\begin{equation}
R_{\rm{S/V}} = \frac{P_{\rm{S}} / I_{\rm{S}}}{p_{\rm{V}} / \tau_{\rm{V}}}
,\end{equation}
where $\tau_{\rm{V}}$ is the optical depth to the quasar, measured in the
V band. We compute this quantity from the extinction $E(B-V)$ as
$\tau_{\rm{V}} = E(B-V) R_{\rm{V}} /1.086$ with $R_{\rm{V}}$, the ratio
of total to selective extinction (e.g., \citealt{PlanckXXI2015}).
We adopt $R_{\rm{V}} = 3.1,$ which is the indicated value for the diffuse
ISM (e.g., \citealt{Fit2004}). Our results are independent of this value,
which only acts as a scaling factor.

While $R_{\rm{P/p}}$ is defined such that it is sensitive to the polarizing
grains alone,
the nondimensional ratio $R_{\rm{S/V}}$ somehow weights $R_{\rm{P/p}}$
by the total amount of dust and its extinction effect on the light of the
background source.
The phenomenology from this ratio is similar to that with $R_{\rm{P/p}}$.
The larger the $R_{\rm{S/V}}$ value, the more prone to dust contamination
the optical polarized signal. Our aim at considering this quantity is to cross
check the results obtained with $R_{\rm{P/p}}$.
For clarity, we base our discussion in the core of the paper on
$R_{\rm{P/p}}$. The results with $R_{\rm{S/V}}$ are given in the Appendix~D.

\medskip

\begin{figure}[t]
        \centering
        \includegraphics[width=\hsize]{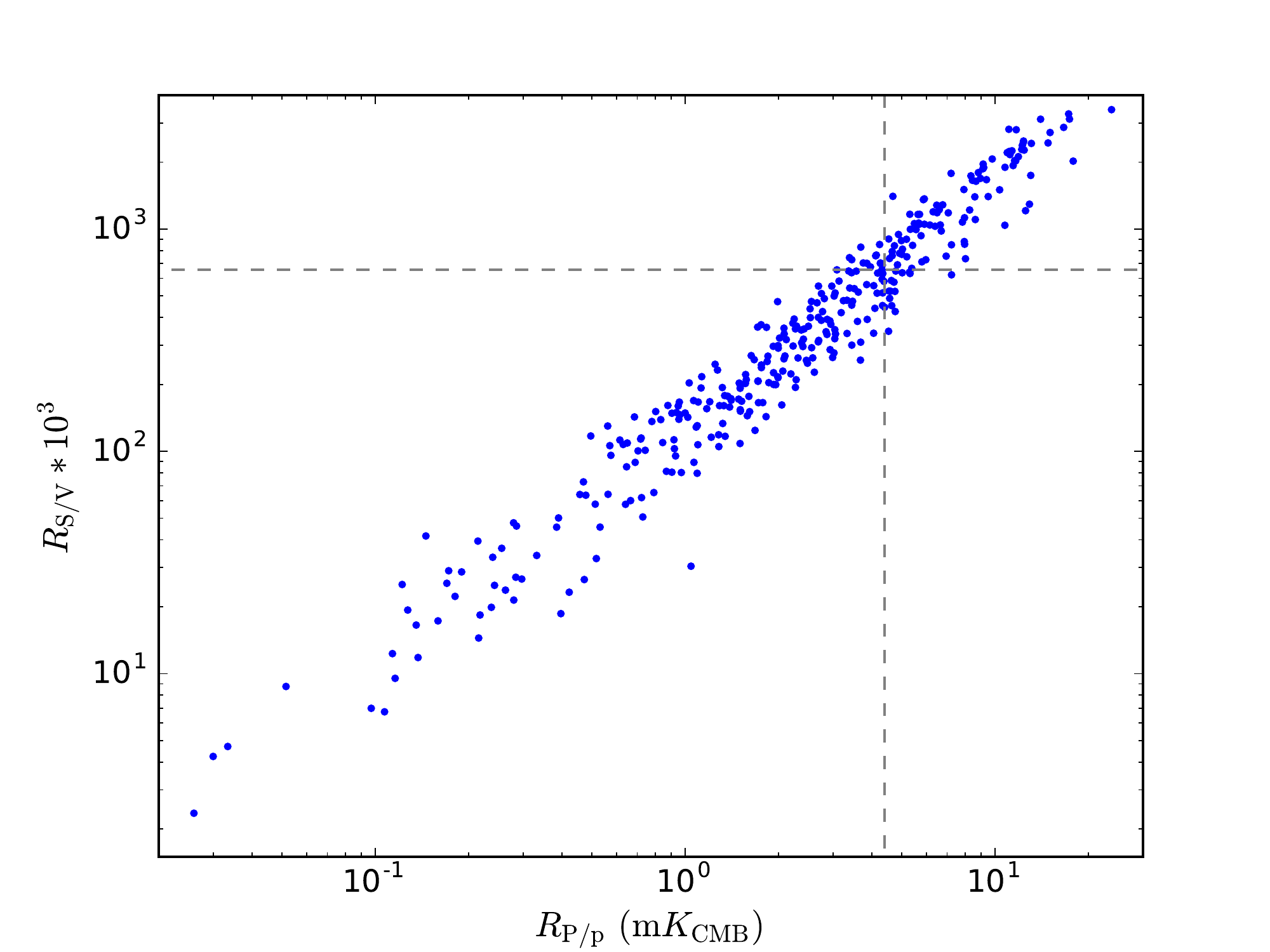}
        \caption{
        Scatter plot in the plane $(R_{\rm{P/p}},\,R_{\rm{S/V}})$ of the 355
        measurements of the polarization ratios measured from the
        \textit{Planck} maps smoothed with a Gaussian kernel of FWHM of 15'
        and the quasar data.
        Horizontal and vertical dashed lines mark the cuts suggested in Sect.~4.1
         to retrieve a $\Delta_{\rm{S/V}}$ distribution that agrees with
        the hypothesis of uniformity at the $2\sigma$ level.
        Our analysis does not allow us to argue for dust contamination for the
        quasars corresponding to polarization ratio values below these
        thresholds.
        }
        \label{fig:RPp-RSV}
\end{figure}

From a phenomenological point of view and if the optical data set is
affected by the Galactic dust at a detectable level, we might expect to
observe a correlation between the angle $\Delta_{\rm{S/V}}$ and the
polarization ratios. The larger the value of the polarization ratios, the
more likely perpendicular the polarization vectors should be.
We search for such correlation in Sect.~3.3.
Fig.~\ref{fig:RPp-RSV} shows the scatter plot of the pairs
$(R_{\rm{P/p}},\,R_{\rm{S/V}})$ for the quasar sample.

\subsection{Preliminary results}
\label{subsec:3.2}
We present the distribution of the 355 $\Delta_{\rm{S/V}}$ measurements
in Fig.~\ref{fig:histDelta}. The blue histogram corresponds to the
$\Delta_{\rm{S/V}}$ that are computed from the position angles of the
quasar optical polarization vectors and the 353 GHz dust polarization
vectors built from the maps smoothed with a Gaussian kernel of
FWHM of 15'. The departure from uniformity is significant and is robust
even accounting for observational uncertainties as illustrated in red in
Fig.~\ref{fig:histDelta} (see Sect.~\ref{subsec:3.3} for the computation
of the uncertainties).
A two-sided Kolmogorov-Smirnov test gives a probability that the blue
histogram is drawn from a parent-uniform distribution of about
$2.4\, 10^{-6}$.
This probability is independent of the number of bins.
Besides, the cumulative-binomial probability to observe, out of 355, 212
or more data points with $\Delta_{\rm{S/V}} > 45^\circ$ is found to be
$P_{\rm{bin}} = 1.5\, 10^{-4}$. This shows that the excess toward the
perpendicularity is significant.
Alternatively, we consider the mean of the cosines of the relative
angles, denoted $\left\langle \cos(\Delta_{\rm{S/V}}) \right\rangle$.
Under the hypothesis that these angles are uniformly distributed in the
range 0 -- 90${}^\circ$, the expected value of this mean is
$2 / \pi \simeq 0.6366$.
Based on Monte Carlo simulations, in which the relative angles are
randomly generated according to a flat distribution, we compute the
probability to observe by chance a value of
$\left\langle \cos(\Delta_{\rm{S/V}}) \right\rangle$ as extreme as that
from the data, the p-value of the observation.
Out of $10^6$ random realizations, none managed to reach the value
obtained for the data, leading to a p-value of
$p_{\rm{value}} \lesssim 10^{-6}$.
This is more than a 5$\sigma$ deviation from uniformity.

We tested the robustness of our results against experimental errors
from both quasar optical polarization data and Galactic thermal dust
polarization data. To account for these uncertainties,
we generated the polarization position angles, polarized intensity, and
degree of linear polarization according to normal distributions centered
on the observations and with their corresponding width. The errors for
the quasars were taken from the catalog and were propagated from the
uncertainties on Stokes obtained by simulations in Appendix A, for the
dust. We produced $10^6$ of realizations.

\medskip

From the sample of 355 quasars, several subsamples were considered
in the literature (e.g., \citeauthor{Hut2005} \citeyear{Hut1998,Hut2005},
\citealt{Jain2004,Pel2014}) to
identify the regions of the parameter space in which the quasars have their
optical polarization vectors more aligned with one another.
For completeness and for comparison with these works, we apply such
historical cuts and test the uniformity of the corresponding
$\Delta_{\rm{S/V}}$ distributions.
We thus separate in two approximately equal parts the sample (\textit{i})
by means of the sky location of the quasars (we separate the northern
to the southern Galactic sky), (\textit{ii}) by means of the degree of linear polarization,
and (\textit{iii}) by means of the redshift $z$ of the quasars.
We also consider combinations of these selection criteria.
Table~\ref{table:Tab_1} summarizes the results using the
$\left\langle \cos(\Delta_{\rm{S/V}}) \right\rangle$ uniformity test. The two
other tests lead to very similar results.

\begin{table}
\caption{Results of uniformity test and historical subsamples.
Probabilities are in per cent. The value $N$ is the number of data points in the
considered subsamples, $p_{\rm{V}}$ the degree of linear optical
polarization (in \%), and $z$ the redshift of the quasars. The
$\left\langle \cos(\Delta_{\rm{S/V}}) \right\rangle$ is measured from
the subsamples and the probabilities $p_{\rm{value}}$ are set using
$10^6$ random simulations of samples of size $N$.}
\label{table:Tab_1}
\centering
\begin{tabular}{l c c c}
\hline\hline
\\[-1.5ex]
Sample  & $N$   &  $\left\langle \cos(\Delta_{\rm{S/V}}) \right\rangle$         & $p_{\rm{value}}$ (\%)   \\
\hline
\\[-1.5ex]
All                     &       355     &       0.552   &       $< 10^{-4}$         \\

$b \geq 30^\circ$
                        &       195     &       0.583   &       $0.80$  \\
$b \leq -30^\circ$
                        &       160     &       0.514   &       $< 10^{-4}$         \\[.5ex]

$p_{\rm{V}} \leq 1.37$
                        &       178     &       0.537   &       $8.0\, 10^{-4}$         \\
$p_{\rm{V}} > 1.37$
                        &       177     &       0.567   &       $0.15$  \\[.5ex]

$b \geq 30^\circ$, $p_{\rm{V}} \leq 1.32$
                        &       98              &       0.551   &       $0.33$         \\
$b \geq 30^\circ$, $p_{\rm{V}} > 1.32$
                        &       97              &       0.573   &       $2.26$         \\[.5ex]
$b \leq -30^\circ$, $p_{\rm{V}} \leq 1.45$
                        &       80              &       0.482   &       $5.0\, 10^{-4}$        \\
$b \leq -30^\circ$, $p_{\rm{V}} > 1.45$
                        &       80              &       0.546   &       $0.45$         \\[.5ex]

$z \leq 1.0$
                        &       178     &       0.504   &       $< 10^{-4}$         \\
$z > 1.0$
                        &       177     &       0.600   &       $5.78$  \\[.5ex]

$b \geq 30^\circ$, $z \leq 0.96$
                        &       98              &       0.511           &       $5.0\, 10^{-3}$        \\
$b \geq 30^\circ$, $z > 0.96$
                        &       97              &       0.655   &       $72.2$         \\[.5ex]
$b \leq -30^\circ$, $z \leq 1.02$
                        &       80              &       0.481   &       $5.0\,10^{-4}$         \\
$b \leq -30^\circ$, $z > 1.02$
                        &       80              &       0.546   &       $0.50$         \\[.5ex]
\hline
\end{tabular}
\end{table}

Under the assumption that the departure from uniformity of the
$\Delta_{\rm{S/V}}$ distribution with an excess toward 90$^\circ$
indicates a contamination of the optical polarization data by Galactic
dust, inspection of Table~\ref{table:Tab_1} reveals that, the
contamination is more noticeable for the part of the quasar sample
with a low degree of linear polarization.
The contamination of the quasar sample could be significant for the
whole southern Galactic sky, even if the departure from uniformity and
the predominance for the perpendicularity appears to be stronger for
the low-redshift part and/or for the low degree of linear polarization
part of the sample.
In the northern Galactic cap, however, the contamination seems to
involve preferentially the low-redshift part of the sample.
The high-redshift part presents even a small deficit, and not an excess,
of polarization vectors at more than 45$^\circ$ to the dust polarization
vectors; as opposed to what contamination would produce. It is
specifically in that region of the space (angular and redshift coordinates)
that the extreme-scale alignments of the quasar optical polarization
vectors have been shown to be the more significant
(\citealt{Hut2005,Pel2014}) in the northern Galactic sky.

The conclusions from our investigation using dust polarization data
agree with those obtained in (e.g., \citealt{Hut2005,Pay2010}) based
on polarized starlight.
All their discussions on the implication of the dust contamination on
the pertinence of the cosmological-scale alignments of the quasar
polarization vectors are thus expected to hold.

\medskip

In Table~\ref{table:Tab_2}, we report the same kind of investigation
but using the median of the polarization ratio $R_{\rm{P/p}}$ to define
the subsamples.
\begin{table}
\caption{Results of uniformity test and polarization-ratio subsamples.
Probabilities are in per cent, $R_{\rm{P/p}}$ in mK$_{\rm{CMB}}$. The value
$N$ is the number of data points in the considered subsample. The
$\left\langle \cos(\Delta_{\rm{S/V}}) \right\rangle$ is measured from the
subsample and the probabilities $p_{\rm{value}}$ are set using $10^6$
random simulations of samples of size $N$.}
\label{table:Tab_2}
\centering
\begin{tabular}{l c c c}
\hline\hline
\\[-1.5ex]
Sample  & $N$   & $\left\langle \cos(\Delta_{\rm{S/V}}) \right\rangle$  & $p_{\rm{value}}$ (\%)   \\
\hline
\\[-1.5ex]
$R_{\rm{P/p}} \leq 2.55$
                        &       178     &       0.610   &       $12.4$          \\
$R_{\rm{P/p}} > 2.55$
                        &       177     &       0.493   &       $< 10^{-4}$         \\[.5ex]

$b \geq 30^\circ$, $R_{\rm{P/p}} \leq 2.46$
                        &       98              &       0.643   &       $58.0$                 \\
$b \geq 30^\circ$, $R_{\rm{P/p}} > 2.46$
                        &       97              &       0.522   &       $0.02$                 \\[.5ex]
$b \leq -30^\circ$, $R_{\rm{P/p}} \leq 2.84$
                        &       80              &       0.581   &       $5.48$                 \\
$b \leq -30^\circ$, $R_{\rm{P/p}} > 2.84$
                        &       80              &       0.447   &       $< 10^{-4}$        \\[.5ex]
%
%
%


\hline
\end{tabular}
\end{table}
Inspection of this table suggests that it is the part of the sample that
corresponds to large values of the polarization ratio from which the
significant departure from uniformity of the $\Delta_{\rm{S/V}}$
distribution toward 90$^\circ$ mainly comes from.
Indeed, cutting the samples in two equal-size subsamples always
leaves a $\Delta_{\rm{S/V}}$ distribution compatible with uniformity
for the low values of the polarization ratio. On the contrary large
values of $R_{\rm{P/p}}$ correspond to significant deviation from
uniformity. This dichotomy is quantified using a two-sample
Kolmogorov-Smirnov test. The probability that the $\Delta_{\rm{S/V}}$
distributions of the low and high $R_{\rm{P/p}}$-value parts are drawn
from the same parent distribution is found to be
$P_{\rm{2KS}}(R_{\rm{P/p}}) = 0.441\%$. We investigate further this
point in the remainder of the paper.

\subsection{Correlation between $\Delta_{\rm{S/V}}$ and $R_{\rm{P/p}}$}
\label{subsec:3.3}
As discussed in Sect.~\ref{subsec:3.1}, $R_{\rm{P/p}}$ (and $R_{\rm{S/V}}$)
represents a good parameter according to which the hypothesis of
dust contamination can be investigated.
In Sect.~\ref{subsec:3.2} we found that $R_{\rm{P/p}}$ could be correlated to the
degree of uniformity of the $\Delta_{\rm{S/V}}$ distribution.
\begin{figure}
        \centering
        \includegraphics[width=\hsize]{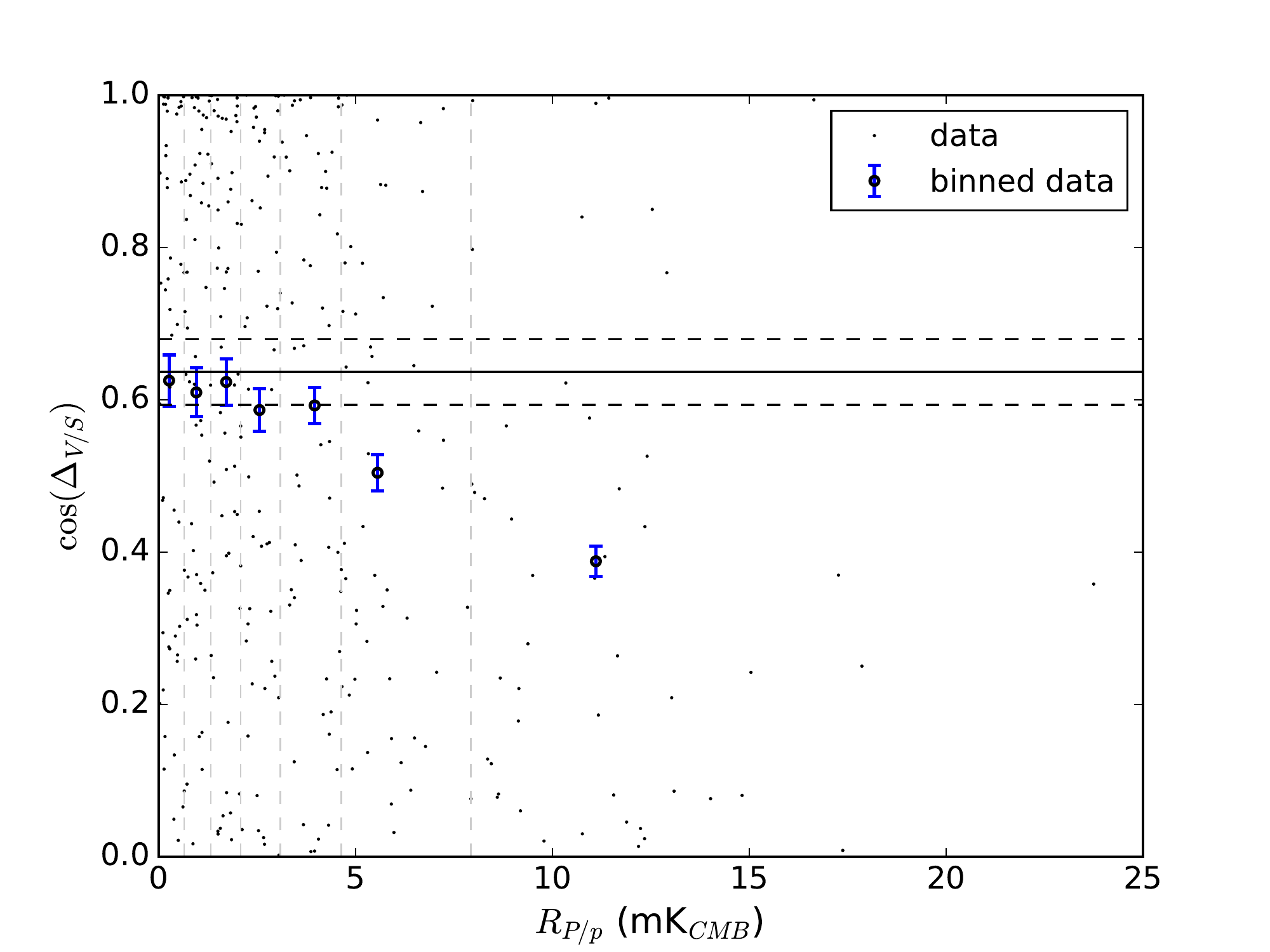}
        \caption{
        Scatter plot in the plane       $(\cos(\Delta_{\rm{S/V}}),\, R_{\rm{P/p}})$
        and binned data. The error bars correspond to the errors on the mean
        of $\cos(\Delta_{\rm{S/V}})$ in bin that are due to experimental
        uncertainties on polarization position angles (dust and quasars).
        Sampling errors in bin are about two times larger.
        Vertical gray lines show the borders of each bins.
        They were chosen such that each bin contains 51 data points, except
        the last one which contains 49.
        The horizontal tick line shows the expected mean under the assumption
        of uniformity in bin and the horizontal dashed lines show the one standard
        deviation from this mean, computed for 51 data points.
        }
        \label{fig:RPpDelta}
\end{figure}
We go further in the investigation of this possible correlation.
In Fig.~\ref{fig:RPpDelta} we show the scatter plot of the couple
$(\cos(\Delta_{\rm{S/V}}),\, R_{\rm{P/p}})$.
In that figure, we also present binned data (blue points).
The adopted sampling is such that each bin contains a fair --and
approximately equal-- amount of data points and in the meantime,
leaves us with as many bins as possible to allow for a detailed
study in terms of the $R_{\rm{P/p}}$ values.

\begin{figure}
        \centering
        \includegraphics[width=\hsize]{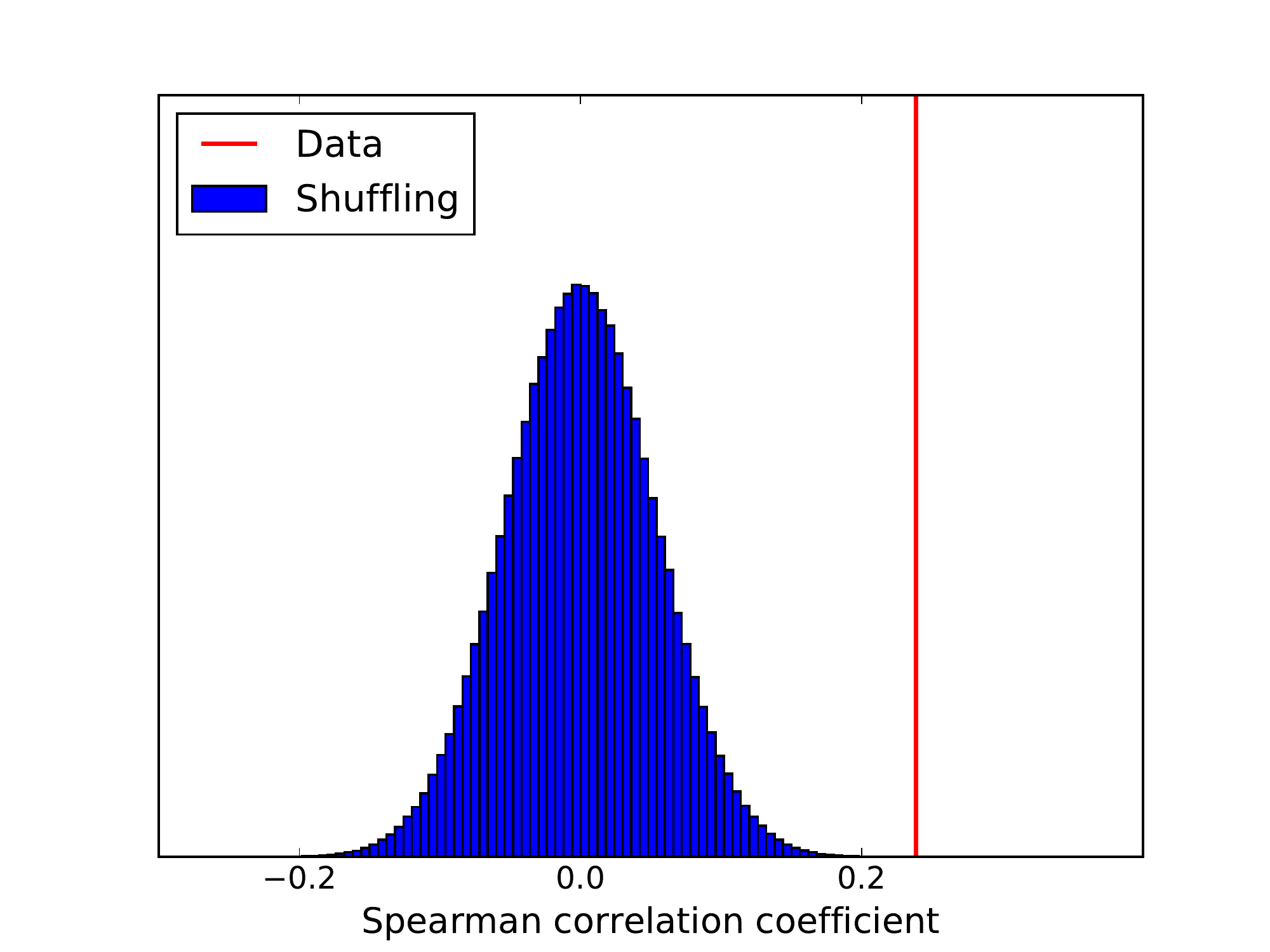}
        \caption{
        Spearman's rank-order correlation coefficients between
        $\Delta_{\rm{S/V}}$ and $R_{\rm{P/p}}$.
        The histogram corresponds to the distribution obtained by shuffling
        $10^6$ time the $R_{\rm{P/p}}$ on the relative angles ($\Delta_{\rm{S/V}}$).
        The (red) vertical line corresponds to the correlation coefficient obtained
        from the data.
        }
        \label{fig:RPpDelta_Spearman}
\end{figure}

As observed in Fig.~\ref{fig:RPpDelta}, the larger the
$R_{\rm{P/p}}$ value, the smaller the
$\left\langle \cos(\Delta_{\rm{S/V}}) \right\rangle$ value.
That is, the larger the $R_{\rm{P/p}}$, the more perpendicular are the
optical polarization vectors with respect to the submillimeter vectors.
We quantify this correlation
using the Spearman's rank-order correlation test. Applied on the pairs
$(\Delta_{\rm{S/V}},\, R_{\rm{P/p}})$, the obtained correlation coefficient
is 0.24 and the two-sided probability of obtaining this result by chance
is $5.5\,10^{-6}$.
We further verify this result by means of a permutation test with $10^6$
random simulations to evaluate the random distribution of the correlation
coefficient. The random realizations were obtained by shuffling the values
of the polarization ratio on the $\Delta_{\rm{S/V}}$ values.
The resulting distribution of the correlation coefficients, corresponding to
the hypothesis of no correlation, is given by the histogram in
Fig.~\ref{fig:RPpDelta_Spearman}.
This distribution has a mean and a
standard deviation of 0.0 and 0.05, respectively. This leads to the conclusion
that the hypothesis of uniformity has to be rejected at 4.5 sigma.
The (one-sided) p-value of the observed correlation, indicated by the (red)
vertical line, is computed as $2.0\,10^{-6}$.

The robustness of the correlation regarding the lines of sight with dust data
having low S/N is discussed in Sect.~\ref{subsec:caveats}.
The robustness of our analysis with respect to the value of the FWHM of
the Gaussian kernel to smooth the dust polarization maps is discussed in
Appendix~B\footnote{We also checked that the deviation from
uniformity of the $\Delta_{\rm{S/V}}$ distribution is also observed if we
consider the 353 GHz $Q$ and $U$ polarization sky maps given by the
COMMANDER component separation algorithm \citet{PlanckX2016}.
These maps have a resolution of 10'.}.
As shown in Appendix~D, this result is also confirmed using the polarization
ratio $R_{\rm{S/V}}$ instead of $R_{\rm{P/p}}$.

\subsection{Other correlations involving $\Delta_{\rm{S/V}}$}
\label{subsec:3.4}
For the sake of completeness, we have searched for similar correlations
to those observed between the $\Delta_{\rm{S/V}}$ and $R_{\rm{P/p}}$
but with the various observables, either from the quasar data, the dust
data, or a mix of those.
These observable are the two polarization ratios $R_{\rm{P/p}}$ and
$R_{\rm{S/V}}$, the dust related quantities $\tau_{\rm{V}}$,
$P_{\rm{S}}$, $p_{\rm{S}}$, $I_{\rm{S}}$ and the quasar quantities $b$,
$l$, $z$, $p_{\rm{V}}$, all defined above.
We led this investigation using the Spearman rank-order correlation test.
The results are given in Table~\ref{table:Tab_3}.
Very significant correlations are detected with the polarization quantities
related to Galactic dust.
\begin{table}
\caption{Correlation between $\Delta_{\rm{S/V}}$ and
the observables.
The polarization ratios $R_{\rm{P/p}}$ and $R_{\rm{S/V}}$,
the dust-related quantities $\tau_{\rm{V}}$, $P_{\rm{S}}$, $p_{\rm{S}}$,
$I_{\rm{S}}$
and the quasar related quantities $b$, $l$, $z$, $p_{\rm{V}}$.
The two-sided probabilities obtained by the Spearman's rank-order
correlation test are reported along with the corresponding correlation
coefficient $\rho$.
}
\label{table:Tab_3}
\centering
\begin{tabular}{l c c}
\hline\hline
\\[-1.5ex]
Observable              & $\rho$        & $P_{\rm{Spe}}$ (\%)   \\
\hline
\\[-1.5ex]
$R_{\rm{P/p}}$  & $0.24$        &       $5.5\,10^{-4}$  \\
$R_{\rm{S/V}}$  & $0.23$        &       $9.1\,10^{-4}$  \\[.5ex]
$\tau_{\rm{V}}$ & $0.13$        &       $1.51$                          \\
$P_{\rm{S}}$            & $0.24$        &       $5.8\,10^{-4}$  \\
$p_{\rm{S}}$            & $0.21$        &       $6.6\,10^{-3}$  \\
$I_{\rm{S}}$            & $0.12$        &       $1.94$                          \\[.5ex]
$b$                                     & $-0.08$       &       $14.6$                          \\
$l$                                     & $-0.03$       &       $61.0$                          \\
$z$                                     & $-0.14$       &       $0.71$                          \\
$p_{\rm{V}}$            & $-0.04$       &       $40.2$                          \\[.5ex]
\hline
\end{tabular}
\end{table}
We also detect a significant correlation between $\Delta_{\rm{S/V}}$
and the redshift of the quasars. This correlation is further discussed
in Sect.~\ref{subsec:4.3}.

\subsection{Corroborating the contamination scenario}
As mentioned in Sect.~3.2, any deviation from uniformity of the distribution
of the $\Delta_{\rm{S/V}}$ angles indicates a possible contamination of the
optical polarization sample by foreground Galactic dust.
However, a nonuniform $\Delta_{\rm{S/V}}$ distribution is not a
self-consistent proof for a contamination. The opposite being also true:
uniformity of $\Delta_{\rm{S/V}}$ distribution does not have to be
considered as a self-consistent proof of no contamination.
Coincidental (un-)correlation of the position angles needs to be investigated.
To that concern, the correlation that we have found between
$\Delta_{\rm{S/V}}$ and the polarization ratio $R_{\rm{P/p}}$ (and
$R_{\rm{S/V}}$) is more convincing as arguments in favor of a measurable
contamination of the quasar optical polarization sample.
Furthermore, the shuffling procedure that we applied in Sect.~\ref{subsec:3.3}
on the pairs $(\Delta_{\rm{S/V}},\,R_{\rm{P/p}})$ and
$(\Delta_{\rm{S/V}},\,R_{\rm{S/V}})$ proves that the departure from
uniformity of the $\Delta_{\rm{S/V}}$ distribution is unlikely accidental
but is rather due to contamination by intervening dust.

\smallskip

To make sure that we are not facing a scenario of fortuitous locations of
lines of sight, we proceed to the two following additional tests.
We rotate the \textit{Planck} dust polarization maps by longitude steps of
five degrees around the north Galactic pole, from $5^\circ$ to $355^\circ$.
For each finite rotation we compute the relative angles between the
quasar and dust polarization vectors, the two polarization ratios and
finally the Spearman correlation coefficients, as for the data.
We further invert the southern and northern hemisphere and proceed
to the same analysis.
This provides us with a distribution of 142 Spearman's correlation
coefficients with which to compare that from the real observations.
Out of the 142 realizations by rotation, none show a correlation
coefficient of the pairs $(\Delta_{\rm{S/V}},\,R_{\rm{P/p}})$ as extreme
as for the data. The same conclusion holds for the pairs
$(\Delta_{\rm{S/V}},\,R_{\rm{S/V}})$.
According to this additional test, the probability that the observed
correlations happen by chance is lower than 1\%.
The Spearman correlation coefficient from the measurements are
$0.02 \pm 0.06$ for the two polarization ratios.
We applied bootstrap resampling method to verify the reliability of the
values of the mean and standard deviation of the distributions.
Utilizing $10^6$ bootstrap realizations from the distribution of the 142
correlation coefficients we found very consistent values. Given the
values of the correlation coefficients for the data, the significance of the
observed correlations is about 3.7 sigma.
A coincidental nonuniformity of the $\Delta_{\rm{S/V}}$ distribution
and of its correlation with polarization ratios are therefore shown as
being very unlikely.

\smallskip

We further rely on an additionally shuffling test to strengthen the
one-to-one correlation that the quasar polarization data seemingly
show with the dust polarization data. We are interested in testing the
hypothesis according to which quasar optical polarization and dust
submillimeter polarization are correlated one another for each line of
sight.
We thus shuffle the quasar data ($p_{\rm{V}}$ and $\psi_{\rm{V}}$)
on the dust data
($P_{\rm{S}}$, $\psi_{\rm{S}}$, ($I_{\rm{S}}$, $\tau_{\rm{V}}$)) to
build the prior distribution of uncorrelated sample.
We proceed to $10^6$ reshuffling. The obtained distribution of the null
hypothesis (no-correlation) has a mean and a standard deviation of 0.006
and 0.05, to be compared to 0.0 and 0.05 obtained in Sect.~\ref{subsec:3.3}.
As above, fortuitous correlation between dust and quasar polarization data
is disfavored given the high significance of the detected correlation ($> 4\sigma$).
We note that according to such a shuffling method, any global systematic
(or bias) or any global offset in the data set should be taken into account
in the computation of the distribution of the null hypothesis.

\medskip

The analyses and results presented in this subsection point toward
the identification of detectable contamination of the quasar optical
polarization data by the intervening Galactic dust.
We made sure that the observed contamination signatures are unlikely to
arise by chance. We are confident in our claim given the significance of the
reported departure from uniformity of $\Delta_{\rm{S/V}}$ distribution, the
correlation between these angles with the polarization ratios, and the
correlation between the polarization of the quasar and dust.

\section{Discussion: mitigating the dust contamination}
\label{sec:discussion}
In this section, we exploit the correlation between the Galactic thermal
dust data and the quasar optical polarization in order to explore the
actual impact of dust contamination on the reported extreme-scale
alignments of quasar optical polarization vectors
without reproducing the cosmological analysis.

\subsection{Search for a new quality criterion}
\label{subsec:4.1}
Figure~\ref{fig:RPpDelta} shows that the departure from uniformity of the
$\Delta_{\rm{S/V}}$ distribution is likely due to the lines of sight having
the largest values of the polarization ratio $R_{\rm{P/p}}$.
It is therefore appealing to introduce a quality criterion of the quasar optical
polarization based on the values of this ratio.
We provide the first investigation in that direction. We show below that this
parameter can be used to remove the data points that are significantly
affected by the dust and therefore obtain a sample of quasar optical
polarization measurements for which the dust contamination is not significant.
Let us emphasize that this does not prevent an actual contamination by
the dust. It rather suggests that the contamination is not strong enough to be
clearly detected. Therefore, in this case, the orientation of the quasar optical
polarization is expected to be not too affected such that we can trust it relates to
the intrinsic polarization.

\medskip

\begin{figure}
        \centering
        \includegraphics[width=\hsize]{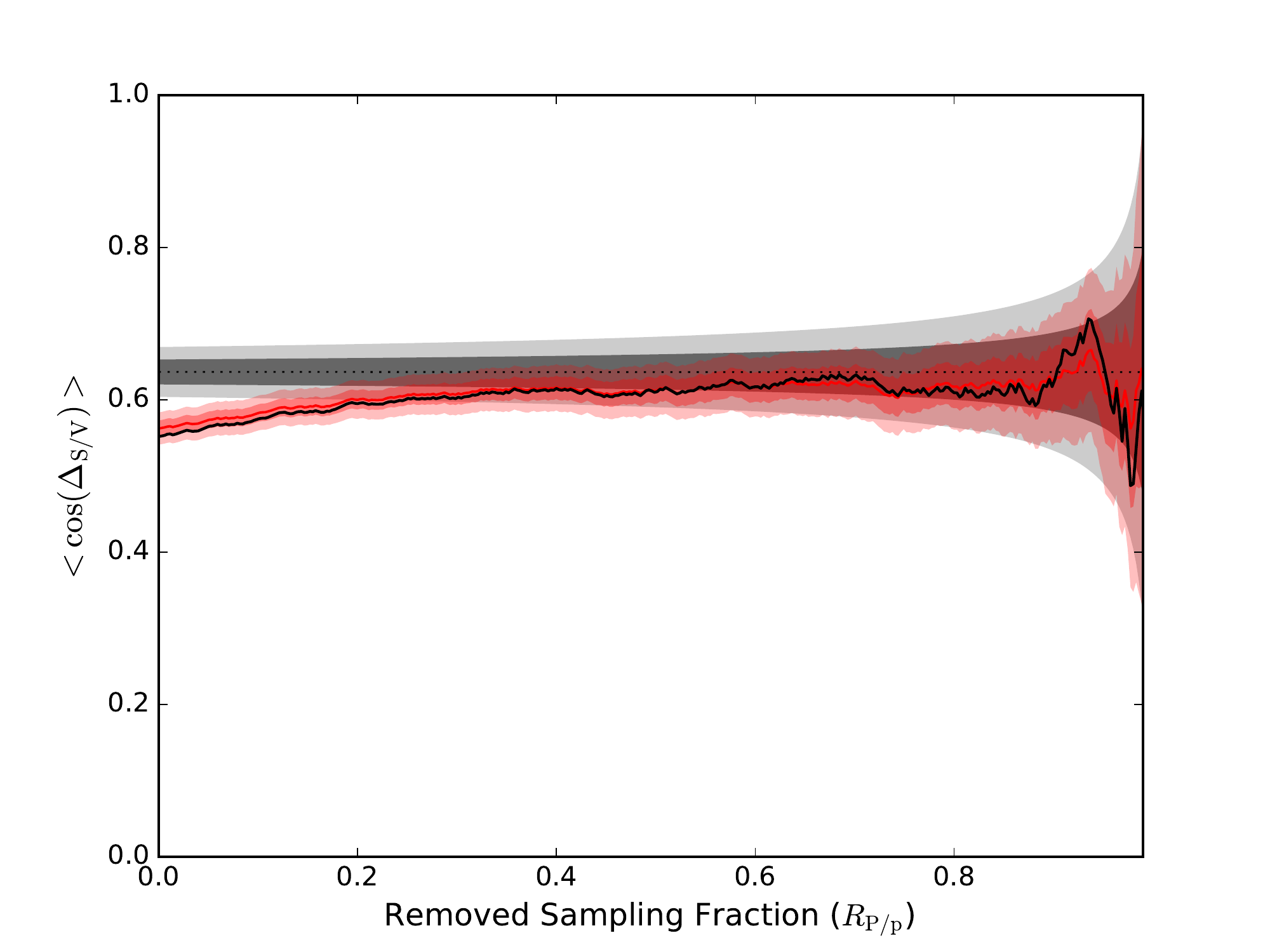} \\
        \includegraphics[width=\hsize]{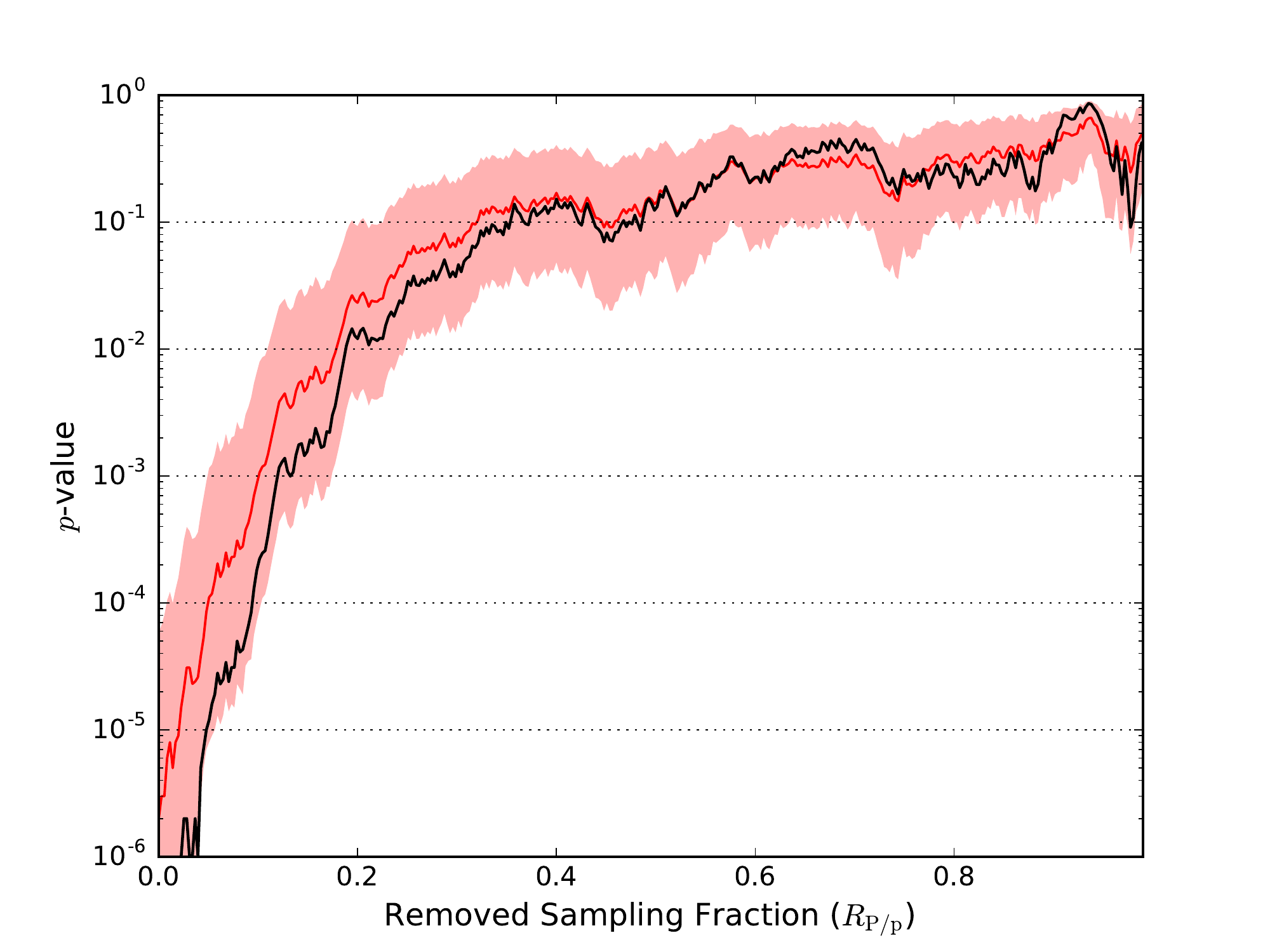}
        \caption{
        (\textit{top})          
        Shown in black is the $\left\langle \cos(\Delta_{\rm{S/V}} \right\rangle$ value
        from the data as a function of the fraction of the sample that is removed
        according to the $R_{\rm{P/p}}$ values of the data points (see text). The red
        line and shaded regions indicate the mean and the one and two sigma contours
        from the observational errors on the polarization position angles (dust and
        quasar).
        The gray shaded regions correspond to the one and two sigma deviations
        around the expected value (black dot horizontal line), assuming a uniform
        distribution of $\Delta_{\rm{S/V}}$ angles.
        (\textit{bottom})
        Under the hypothesis of uniform distribution of $\Delta_{\rm{S/V}}$, the black
        line is the p-value of the observation as a function of the removed fraction of
        the sample. The red line and shaded region correspond to the p-values
        computed for the mean and the one standard deviation from the mean
        of the observation accounting for the position angles uncertainties.
        }
        \label{fig:SamplingFraction_RPp}
\end{figure}
In Fig.~\ref{fig:SamplingFraction_RPp} (\textit{top}),
starting with the original sample of 355 $\Delta_{\rm{S/V}}$
measurements, we gradually reduce the sample size by removing
those data points having largest value of the polarization ratio $R_{\rm{P/p}}$.
It is shown that gradually removing the part of the sample corresponding to
the higher values of the $R_{\rm{P/p}}$ ratio leads to
$\left\langle \cos(\Delta_{\rm{S/V}}) \right\rangle$ values more in agreement
with the hypothesis of a uniform distribution.
This is illustrated in Fig.~\ref{fig:SamplingFraction_RPp} (\textit{bottom})
where we show the one-sided p-value of the observation for each removed
sampling fraction. These are the probabilities that a sample drawn from a
parent uniform distribution leads to a value larger than that of the observed
$\left\langle \cos(\Delta_{\rm{S/V}}) \right\rangle$ of the truncated sample.
The one-sided p-values of the sample corresponding to truncation of 20, 30,
and 40\% are respectively, 1.2\%, 4.6\%, and 15.2\%.
Removing about 30 per cent of the sample is sufficient to retrieve a
$\cos(\Delta_{\rm{S/V}})$ distribution for which the hypothesis of uniformity
cannot be rejected at the two sigma level.

\begin{figure}
        \centering
        \includegraphics[width=\hsize]{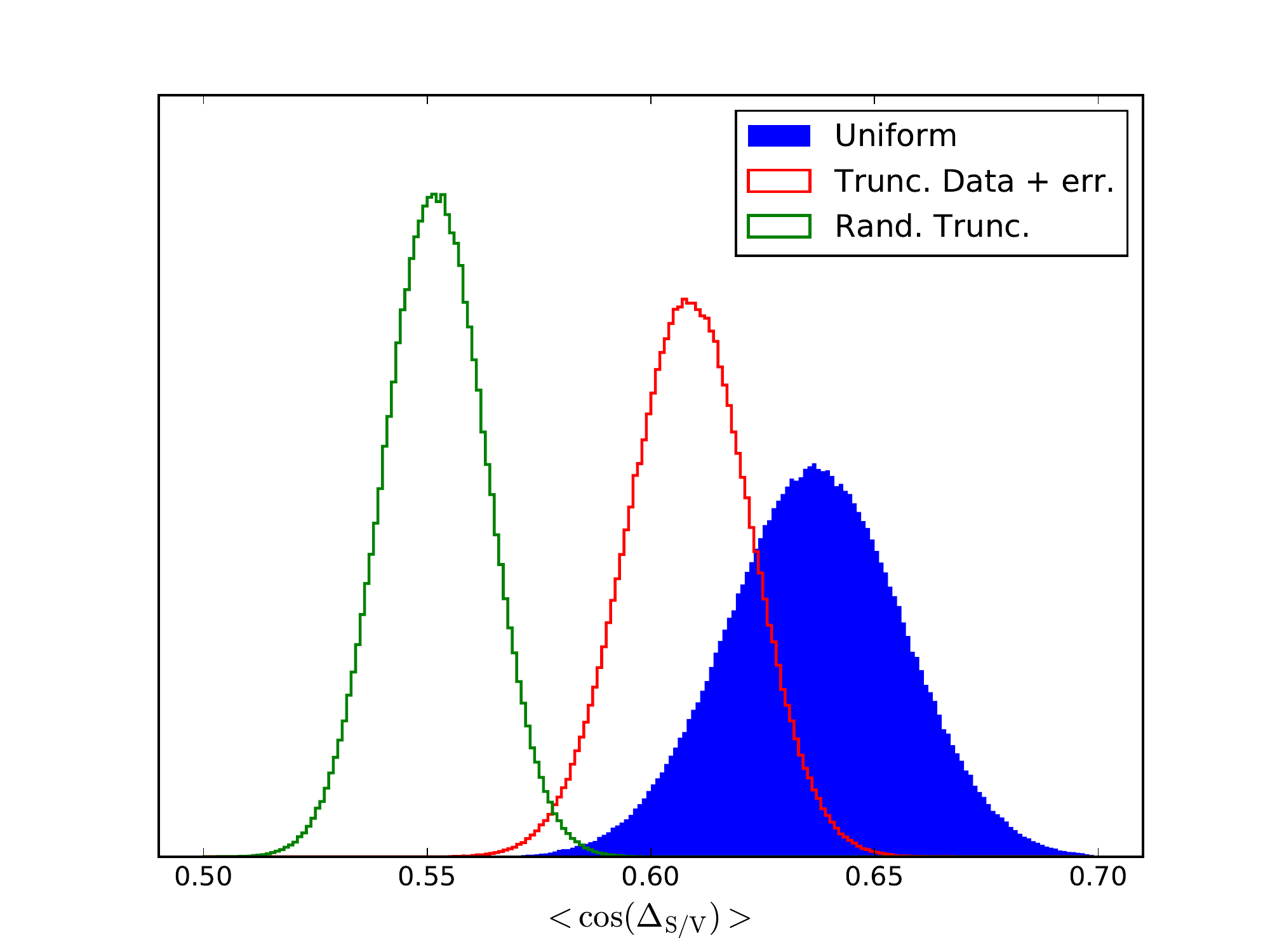}
        \caption{
        Filled blue histogram shows distribution of
        $\left\langle \cos(\Delta_{\rm{S/V}}) \right\rangle$ for $10^6$ realizations of uniform
        $\Delta_{\rm{S/V}}$ distributions for a sample size of $0.7 \times 355$.
        The red unfilled histogram corresponds to the values obtained
        for the data when removing the 30\% of the sample with the higher
        values of the $R_{\rm{P/p}}$ ratio. This distribution takes into account
        the errors on the polarization position angles.
        The green unfilled histogram shows $10^6$ realizations of 30\% random
        truncation of the original sample.
        }
        \label{fig:TruncationRPp}
\end{figure}
We further show, in Fig.~\ref{fig:TruncationRPp}, that the polarization
ratio is a good parameter at discriminating quasar optical polarization
data that are strongly contaminated by dust.
To that aim, we compare the $\left\langle \cos(\Delta_{\rm{S/V}})) \right\rangle$ obtained after
removing the 30\% corresponding to the largest $R_{\rm{P/p}}$ values from the sample to
that when randomly removing 30\% of the sample.
It is shown that
(\textit{i})
the $\left\langle \cos(\Delta_{\rm{S/V}}) \right\rangle$ values obtained by
random truncation of the sample keep deviating significantly from the
hypothesis of uniformity,
(\textit{ii})
the values of $\left\langle \cos(\Delta_{\rm{S/V}}) \right\rangle$
corresponding to the sample truncated by means of $R_{\rm{P/p}}$
do not correspond to a random truncation of the sample, and finally
(\textit{iii})
that the latter values do not violate the hypothesis of a uniform distribution
of the $\Delta_{\rm{S/V}}$ angles at the two sigma level.
In Appendix~D, we show that similar result is obtained for $R_{\rm{S/V}}$
and also for the other values of the FWHM of the Gaussian kernel used
to smooth the dust maps.

In order to minimize the impact from dust contamination in the quasar
sample, it appears relevant to discriminate the data points with respect
to their value of the polarization ratios.
Our analysis suggests removing at least all data points with either
$R_{\rm{P/p}} \geq 4.4\, {\rm{mK}}_{\rm{CMB}}$ or
$R_{\rm{S/V}} \geq 6.5\,10^{-3}$ from the original sample.
However these values depend on the width of the smoothing
kernel used to compute the polarization ratios from the polarized maps.
These threshold values have to be regarded with caution. First
because they are derived from the sample itself and second, because
the remaining fraction of the sample contains the lines of sight for which
the dust polarization quantities are poorly known owing to the
\textit{Planck} sensitivity and the very diffuse character of the ISM (see
below).

\begin{figure*}[t]
        \centering
        \begin{tabular}{cccc}
        \includegraphics[trim={1.1cm 0cm 2.cm 0cm},clip,width=.23\linewidth]{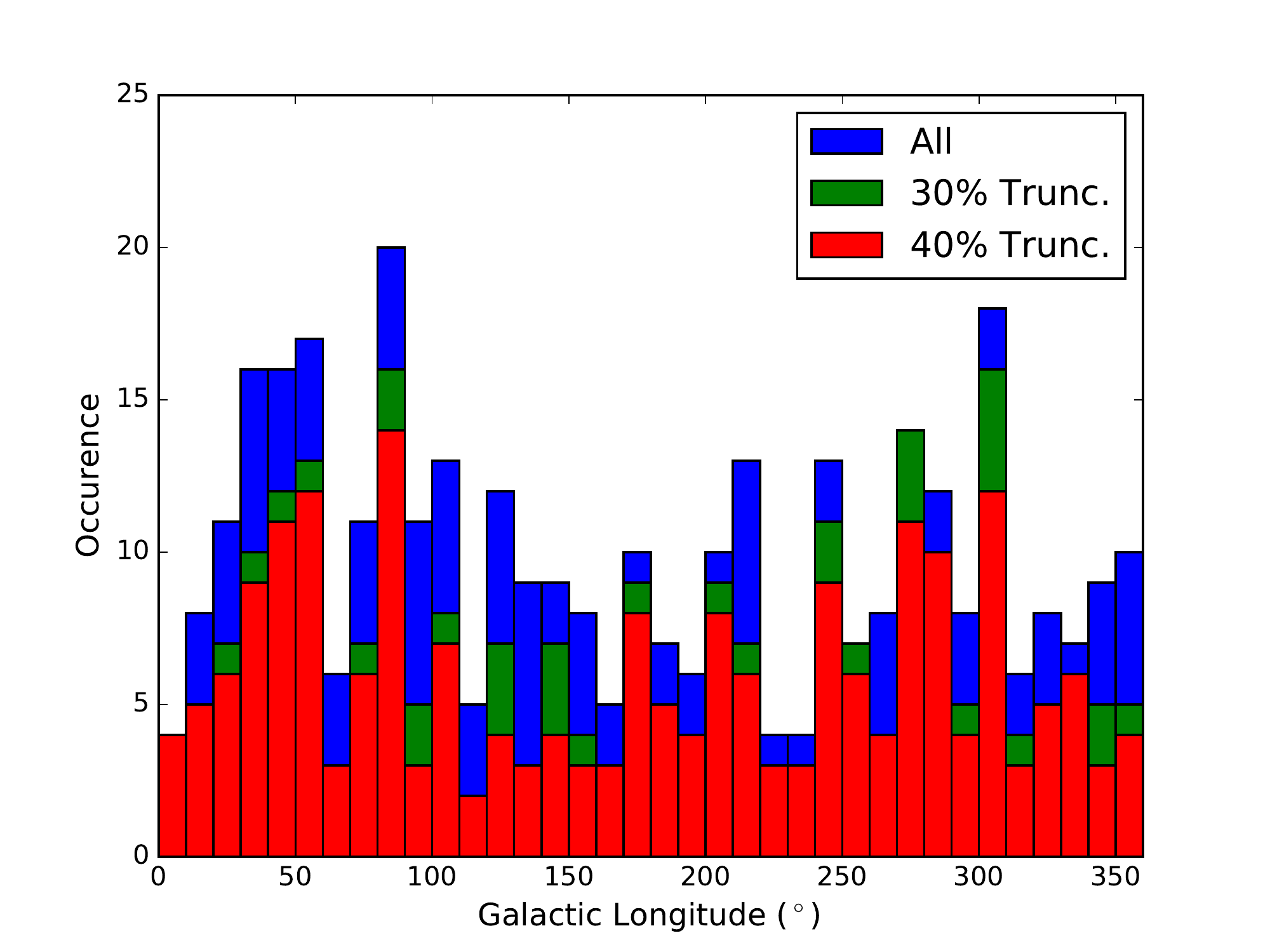} &
        \includegraphics[trim={1.1cm 0cm 2.cm 0cm},clip,width=.23\linewidth]{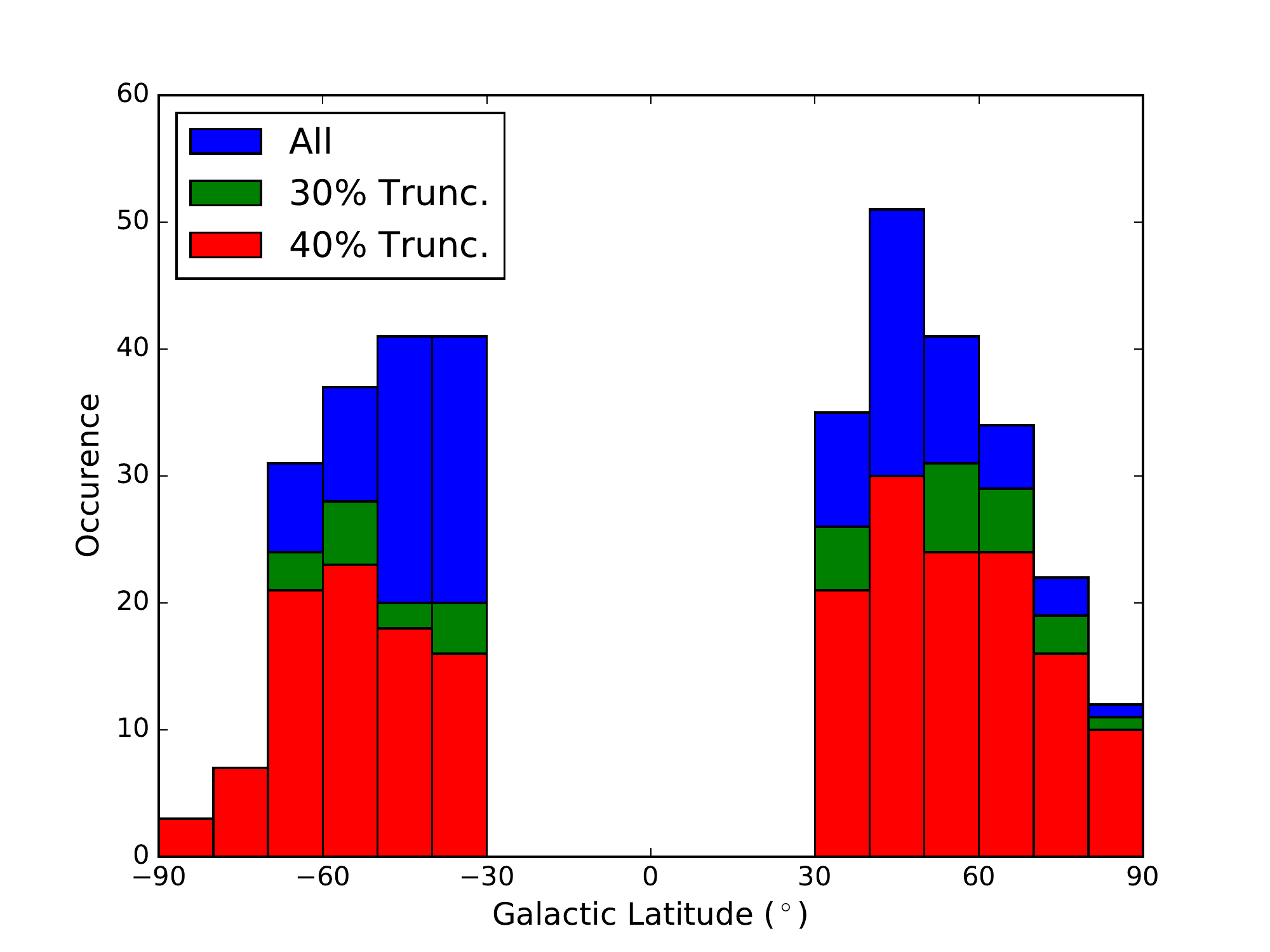} &
        \includegraphics[trim={1.1cm 0cm 2.cm 0cm},clip,width=.23\linewidth]{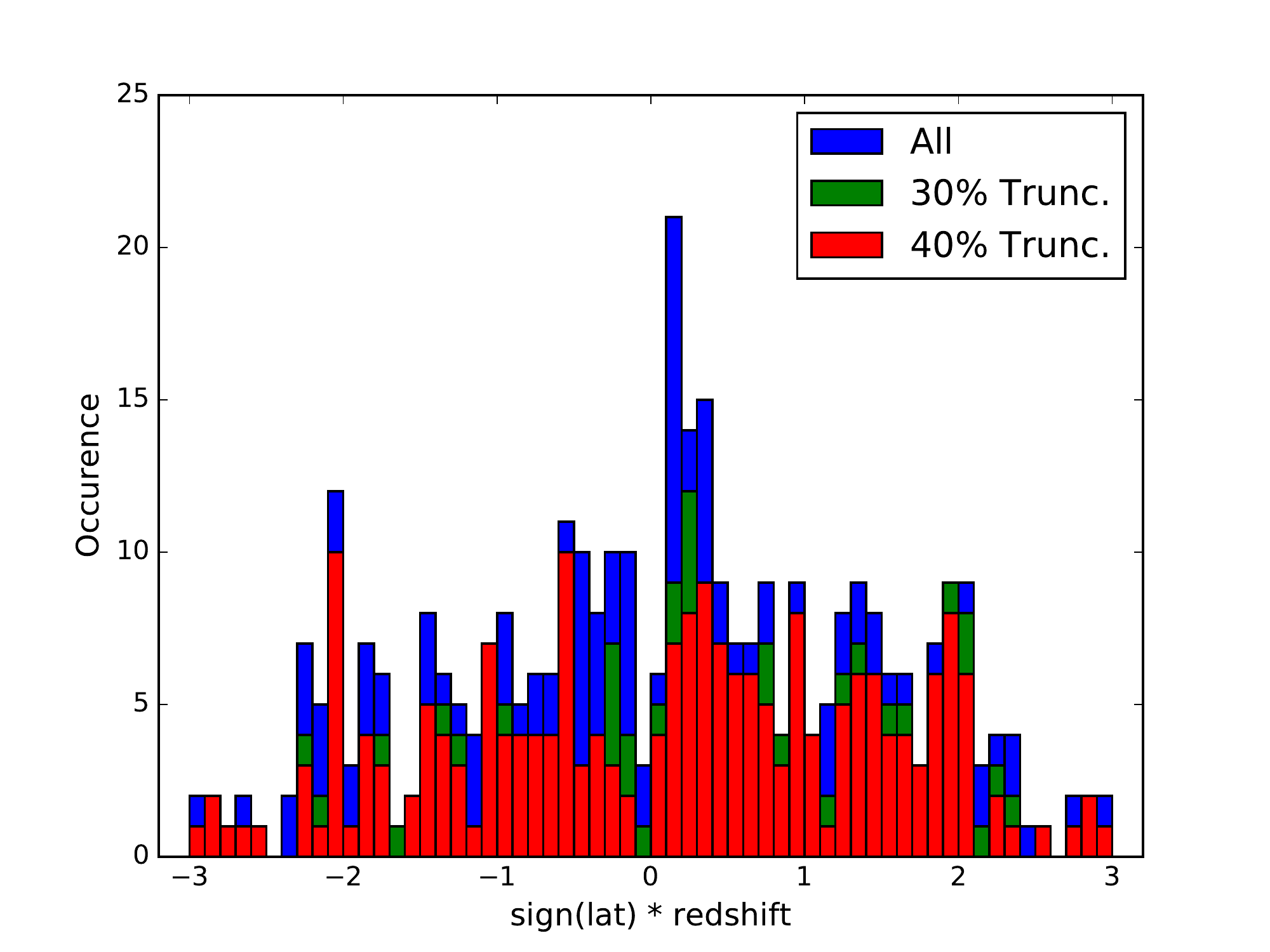} &
        \includegraphics[trim={1.1cm 0cm 2.cm 0cm},clip,width=.23\linewidth]{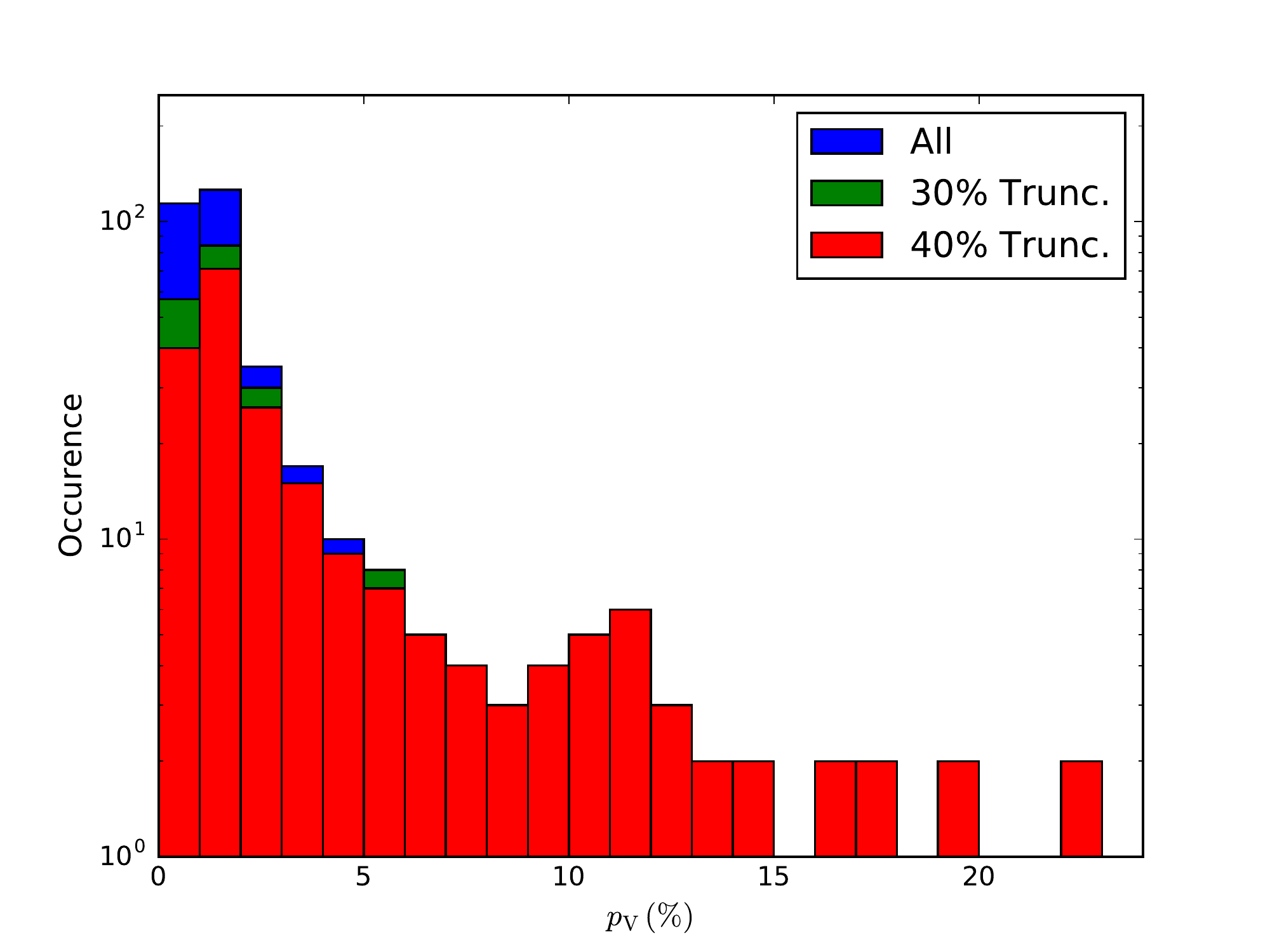}     \\         
        \end{tabular}
        \caption{
        Distribution of the parameters attached to the quasars without and
        with selections by means of the $R_{\rm{P/p}}$ ratio. From top left
        to bottom right are Galactic longitude, Galactic latitude,  signed
        resdhift (to differentiate the northern ($>0$) and southern Galactic
        cap ($<0$)) and degree of linear polarization.
        The filled blue histogram represents the full catalog, green and red
        show the 30 and 40\% truncation, respectively. We note the log-scale
        for the last plot.
        }
        \label{fig:TruncQSOhist}
\end{figure*}
\subsection{Caveats and mitigation}
\label{subsec:caveats}
A question arises about how and how much our results are sensitive to
the part of the sample that corresponds to the lines of sight for which
only low S/N data are available for the
Galactic thermal dust polarized emission. We address this question below
in investigating the impact of the polarization bias on the significance of
the reported correlation. We further infer the impact of the low S/N data
points on the suggested quality criteria to be imposed to the quasar
sample. We finally provide physical arguments to not discard those data
points with low S/N dust polarization data.

\subsubsection{Robustness of the correlation and low S/N data}
By construction, noisy polarization data are subject to a positive bias that
artificially produces larger values of the polarized intensity
\citep[e.g.,][]{War1974}.
While the quasar optical polarization data have been corrected for
this bias in the catalog, we did not attempt to correct the dust
polarization data for it.

First, let us note that $\sigma_{\rm{P_{\rm{S}}}}$ is largely uniform
at high Galactic latitudes owing to the scan strategy of \textit{Planck}.
As a consequence, in our sample, the S/N depends quasi-linearly
on the $P_{\rm{S}}$ values, which implies that the low S/N of the
polarized intensities correspond to low $P_{\rm{S}}$ values which,
in turn, are more prone to the polarization bias.
As a result, within the sample from which we reported a correlation
between $R_{\rm{P/p}}$ and $\Delta_{\rm{S/V}}$ (see
Sect.~\ref{subsec:3.3}), it is the part with the low $R_{\rm{P/p}}$
values that is the more likely affected by the polarization bias.
Because the polarization bias leads to artificially increased values
of $P_{\rm{S}}$ and thus of $R_{\rm{P/p}}$ values, the bias can only
act to reduce the significance of the correlation.
Indeed, in principle the polarization bias affects only points to
the left of Fig.~3 (low $P_{\rm{S}}$ values), by shifting these points to
the right. The correction of the bias would therefore in principle lead to
a small shift of these points to the left. Correcting for polarization bias
would therefore result in a tighter correlation.

Therefore, keeping the data points with low S/N mitigates the detection
of the dust contamination, which is already significant.
We thus found unnecessary to correct for this effect using, for example,
debiasing methods such as that presented by \cite{Pla2014}.
Meanwhile, we had repeated most of the analyses but removed the sight
lines for which $P_{\rm{S}} < \sigma_{\rm{P_S}}$. With that reduced
sample we found consistent results but with lower significance. For instance,
the probability given by the Spearman correlation test that the correlation
between $\Delta_{\rm{S/V}}$ and $R_{\rm{P/p}}$ is due to chance was found
to be $1.0 \, 10^{-4}$ instead of $5.5 \, 10^{-6}$. This is the opposite of
what we expect from the above. This is because we are dominated by
the sampling variance rather than by the observational uncertainties.

\subsubsection{Quality criterion and low S/N data}
While investigating the fraction of the sample that has to be removed to
define a quasar optical polarization sample where the possible
dust contamination is minimized,
we have to worry about the role of the sight lines with low S/N dust data.
It is clear that those data points likely contribute to the uniformity of the
$\Delta_{\rm{S/V}}$ distribution.
This is because large uncertainties on the dust polarization position angles
would act to smear the correlation between dust and quasars, if there is
any. We can decide to reject those lines of sight from our analysis.
If we reject all lines of sight with $P_{\rm{S}} < \sigma_{\rm{P_S}}$ and
reproduce the truncation analysis presented in Sect.~\ref{subsec:4.1},
we find that the $\Delta_{\rm{S/V}}$
distributions keep deviating from uniformity, with a p-value as high as
1 -- 5 \%, even for large values of the removed fraction.
This does not implies that a fraction of the corresponding quasar
sample cannot be used for cosmological inference.
Indeed, for low $R_{\rm{P/p}}$ values, irrespective of the noise level of
the dust polarization data, the contamination by dust of the quasar optical
polarization data might be too inefficient to smear the intrinsic
quasar optical polarization.

\medskip

As a corollary, the above discussion shows that we need the sight lines
for which the dust data are dominated by the noise to retrieve a
distribution of the $\Delta_{\rm{S/V}}$ that is compatible with uniformity,
i.e., that the contamination is only marginally detected based on the
relative angle criterion. It is thus legitimate to attempt to figure out if this
uniformity is physical or due to lacking quality of the dust polarization
data. This point is addressed below.

\subsubsection{Low S/N and inefficient contamination}
As already mentioned, low S/N dust polarization data corresponds to low
$P_{\rm{S}}$ in our sample.
Based on a phenomenological model of the polarized dust
emission
\cite[see, e.g.,][for a review and similar discussion]{PMR2018,PM2018},
we know that low $P_{\rm{S}}$ values correspond either to lines of sight
with low dust column density or to lines of sight along which depolarization
is at play because of conspiring geometry of the Galactic magnetic field. Indeed,
the Galactic magnetic field could be such that the observed polarized
intensity is low because of the integration of many different preferred
directions along the line of sight. In the former case low total intensity of
the dust is also to be expected while in the latter case the dust intensity
are expected to be large as resulting from the integration of the dust density
along the line of sight. We note that in case of depolarization, even if no
particular correlation would be observed between the orientations of the
submillimeter and optical polarization vectors, we could not safely consider
the measured quasar optical polarization as being intrinsic to the sources.
In such a case our analysis would become senseless.

To verify that we are not facing such a scenario, we consider those lines
of sight for which the S/N of the dust polarized signal is low.
We observed that they correspond preferentially to low $I_{\rm{S}}$
values, corresponding to low column density. This has been tested by
means of a two-sample Kolmogorov--Smirnov test that leads to a
probability below $3.0\,10^{-4}$ that the $I_{\rm{S}}$ distribution of the
subsample is a random subset from that of the whole sample.
This situation is even more pronounced by means of the optical depth
$\tau_{\rm{V}}$.
This points toward a scenario according to which there is not enough
dust along the sight lines to significantly modify the polarization state of
the optical light from the background quasars. Therefore, the observed
uniformity of the $\Delta_{\rm{S/V}}$ distribution of this subsample of low
S/N dust data is seemingly physical.
Relying on this argument, we shall prefer to not reject that part of the
sample from our analysis.

\medskip

Even if the above does not allow us to thoroughly conclude for no-dust
contamination of the subsample of the quasar optical polarization
corresponding to the sight lines with low S/N, we have to acknowledge
the arguments pointing toward a low contribution of the Galactic dust
on those polarization data.

\subsection{Do the extreme-scale alignments survive?}
The result obtained in the previous subsections show that, based on our
contamination criterion, Galactic ISM contamination is not detected
at the two sigma level
for about 70 per cent of the quasar optical polarization sample presented
by \citet{Hut2005}.
We argue that this fraction of the sample could be used for cosmological
investigations as we expect the impact of the dust contamination on the
cosmological analysis to be small.
It is beyond the scope of the present paper to reproduce the study of the
large-scale anisotropies of the quasar optical polarization vector orientations.
Nevertheless let us present some hints that points toward the fact that the
extreme-scale alignments cannot fully be explained by a dust contamination
scenario.

\subsubsection{Redshift dependence}
\label{subsec:4.3}
In Sect.~\ref{subsec:3.4} we reported on a significant correlation between
$\Delta_{\rm{S/V}}$ and the redshift of the quasars. This actually reflects
the redshift dependence of the preferred orientation of the quasar optical
polarization vectors (at least in the northern Galactic hemisphere). Indeed,
in the northern hemisphere, quasars seem to have --on average-- their
polarization vectors more randomly oriented with respect to the dust (second
line of Table~\ref{table:Tab_1}).
Interestingly, low-redshift quasars have their polarization vector
preferentially perpendicular to the dust polarization vectors.
The tendency is at the opposite for high-redshift quasars. This dichotomy
is well observed by comparing the 12th and the 13th lines of
Table~\ref{table:Tab_1}.

In the past, \citeauthor{Hut2005} (\citeyear{Hut1998, Hut2001, Hut2005})
reached compatible conclusions considering starlight polarization.
This observational fact provided an actual strong argument against
the hypothesis that dust contamination would be at the origin of the
reported extreme-scale alignments of the quasar optical polarization
vectors.
The reason is that the correction of a strong Galactic contamination
at low redshift would automatically imply much stronger alignments
at high redshift as the lines of sight of the two regions pass through
the same Galactic space \citep{Pay2010}.
Therefore, in at least one of the two gigaparsec-scale regions of the space
defined by Hustem{\'e}kers et al. toward the north Galactic pole,
the quasars are expected to have their optical polarization vectors
effectively aligned one another.
This argument holds as far as the lines of sight to those quasars
indeed probe the same ISM. This was not an a priori guarantee given the,
sometimes rapid, spatial variation (on sky) of the column density of
the polarizing dust and the induced polarization direction.

In this work, by highlighting a redshift dependence of the preferred orientation of
the quasar optical polarization vectors directly compared to the dust polarization
vectors, we provide further credits to the old arguments, discarding a
scenario of conspiring dust features.

\subsubsection{Alignment regions and strong contamination}
From previous works (\citealt{Hut1998,Hut2005,Pel2014}),
we know that the alignments of the polarization vectors are stronger, i.e.,
more significant, in well-defined regions of the three-dimensional space.
We have checked that the quality criteria based on polarization ratios
does not favor a depopulation of those regions, compared to other places.
For completeness, we give in Fig.~\ref{fig:TruncQSOhist} the histograms
of the physical parameters of the quasars belonging to the whole catalog
and with 30 and 40 per cent truncation according to the $R_{\rm{P/p}}$ values.
It can be seen that the selection occurs preferentially in some regions
of the parameter space\footnote{We note that this was already known from
studies relying on the starlight optical polarization inquiries \citep{Hut2005}.}.
However, these are not those in which the Gpc-scale alignments were
reported as being the more significant. For example, the so-called
A1 region was defined according to the limits \citep[see, e.g.,][]{Hut2005}:
$168^\circ \leq$ RA $\leq 218^\circ$,  DEC $\leq 50^\circ$ and
$1.0  \leq z \leq 2.3,$
where RA and DEC stand for the right ascension and declination of
the quasars. In the whole sample, this region
contains 56 quasars. Among those, 46 (38) pass the selection of the 30
(40) per cent truncation based on $R_{\rm{P/p}}$ values, to be compared
to the expected 39.1 (33.6).

Based on the above, we cannot argue for significant changes in the
relevance of the extreme-scale alignments of the quasar optical polarization
vectors because the alignments were discovered in a sample having half of
its current size and we do not find stronger contamination by intervening
Galactic dust of the quasars belonging to the regions where the alignments
are known to be the more significant. To the question of whether or not the
extreme-scale alignments of the quasar optical polarization vectors
survive to the Galactic dust contamination scenario, the answer is likely
yes. A clear demonstration of this claim would require further investigation
relying on a large number of simulations and on the elaboration of  a
framework within which the contamination by intervening Galactic dust
would be reliably modeled
and removed from the quasar data.

\section{Conclusions and perspectives}
\citeauthor{Hut2005} (\citeyear{Hut1998, Hut2001, Hut2005}) have
reported alignments of quasar optical linear polarization vectors
extending over cosmological-scale
regions of the Universe. 
Severe quality criteria, which considerably limit the sample size, were
adopted to reduce at best the Galactic contamination of the quasar
optical polarization data by dust.
The authors concluded that even if the resulting sample is unavoidably
contaminated to some level by the ISM, this contamination would never
be able to account fully for the observed characteristics of the
extreme-scale alignments.
The redshift dependence of the alignment directions is one of the
characteristics of the alignments that cannot be easily explained with
Galactic foregrounds (see \citealt{Pay2010} and also our Sect.~4.3.1).

In this paper, we reinvestigate the degree of contamination of these
optical polarization data by Galactic dust using \textit{Planck} data
instead of starlight optical polarization measurements.
Our goal was to
make sure that the contamination level of the quasar optical polarization
sample is low and to investigate the possibility that complex Galactic
dust features could account for the alignments.
We led our investigation using the \textit{Planck} full-sky maps at 353
GHz, which capture the diffuse thermal dust polarized emission. This
observable is complementary and independent from starlight-based
polarization estimation of the ISM. It has the advantage to
have an homogeneous sky sampling and to be representative of the
whole integration of the dust along the line of sight.

\medskip

First, our analysis reveals a significant deviation ($> 5\sigma$) from
uniformity of the distribution of the relative angles between the
polarization vectors of the quasars and the dust ($\Delta_{\rm{S/V}}$).
There is a significant tendency toward the perpendicularity of those
vectors.
Second, we found significant correlations
($\sim 4\sigma$) between the relatives angles and two polarization
ratios that are computed for each line of sight to a quasar.
The larger the values of the polarization ratios ($R_{\rm{P/p}}$ or
$R_{\rm{S/V}}$), the more perpendicular the optical polarization vectors
to the dust polarization vectors.
Therefore, using dust maps, we were able to diagnose a stronger contamination
of the quasar optical polarization data by the Galactic dust than was
previously thought (\citeauthor{Hut1998} \citeyear{Hut1998,Hut2001,Hut2005}).
Indeed, relying on criteria based on the distribution of
the relative angles and on two polarization ratios, we revealed a measurable
effect of the intervening Galactic dust on the quasar optical polarization
data. We further showed that, based on the relative angle criterion, the
contamination is only marginally detected when removing at least roughly
30 per cent of the sight lines with high values of the polarization ratios.
This suggests that for the remaining lines of sight the contamination by the
intervening Galactic dust is likely inefficient at smearing the intrinsic optical
polarization of the quasars. This reduced sample could then be used for
cosmological inference. We thus provide a new quality criterion based on the
polarization ratio values to minimize at best the ISM contamination.

While a definitive proof requires further developments and analyses, we
found arguments that reinforce previous claims according to which the
ISM cannot be responsible for the cosmological-scale alignments of the
quasar optical polarization vectors.
The cosmological relevance of the extreme-scale alignments of the quasar
polarization vectors is thus expected to be unchanged as also discussed at
the end of Sect.~4.3. We further reinforced the argument against a Galactic
contamination for the quasars that have high redshift and that are located in
the northern Galactic cap. The tendency of the $\Delta_{\rm{S/V}}$ distribution
is indeed opposed to what a contamination would produce.

\medskip

Further inquiries would require reproducing the analyses presented
in \citet{Hut2005} and \citet{Pel2014} with the restricted sample, i.e.,
applying the polarization-ratio selection.
Alternatively, providing a fair modeling of Galactic dust contamination
of the optical polarization data, a decontamination of the sample should
be investigated. The observational characteristics of the extreme-scale
alignments of the quasar polarization vectors shall be adapted
accordingly. While this is well beyond the purpose of this work, this
task is mandatory in a search for the physical scenario responsible
for these striking observations.
For such an inquiry, additional quasar optical polarization data will be
required.
Quasars showing a low degree of linear polarization should particularly
be included in the sample as they could provide upper limits to the dust
contribution of the optical polarized signal.
Together with polarization data of Galactic stars, they could be used to
fit the parameters of the contamination model.
Unfortunately, and to the best of our knowledge, there is yet no publicly
available large quasar optical polarization sample of that sort.
Hopefully this work would motivate the elaboration of such a sample.

\medskip

Large-scale correlations of the quasar polarization vectors have also
been observed at radio wavelengths (8.4 GHz) \citep{Pel2015} in a
large sample of flat-spectrum radio sources compiled by \cite{Jac2007}.
While there were arguments against the hypothesis of strong Galactic
contamination, it should be worthwhile to perform a comparison similar to
that presented in this work, but between the synchrotron full-sky map
released by \textit{Planck} and the point-like radio source
polarization vectors. To this regard, the polarization maps of the diffuse
emission and the point-like source catalog at low frequencies (11 -- 20
GHz) that will be released by the
{\small{RADIOFOREGROUNDS}}\footnote{http://www.radioforegrounds.eu}
project using data from the {\small{QUIJOTE}} experiment
\citep{QUIJOTE2017} will constitute a considerable advantage to lead
such inquiries in the northern equatorial hemisphere.

\begin{acknowledgements}
        I thank C{\'e}line Combet, Damien Hutsem{\'e}kers, Juan
        Mac{\'i}as-P{\'e}rez, Florian Ruppin, and Nicolas Ponthieu for their
        generous advice and the inspiring and guiding discussions that led
        to a significant improvement of this manuscript. I also thank Vincent
        Guillet for a late discussion and for pointing out a unit mistake to me.
        I acknowledge my referee R. Genova-Santos for his help in clarifying
        the discussion and in highlighting the conclusions of this work.
        This work has been partially funded by the European Union' s
        Horizon 2020 research and innovation program under grant
        agreement number 687312.
        We acknowledge the use of data from the Planck/ESA mission,
        downloaded from the Planck Legacy Archive, and of the Legacy
        Archive for Microwave Background Data Analysis (LAMBDA).
        Support for LAMBDA is provided by the NASA Office of Space
        Science. Some of the results in this paper have been derived using
        the HEALPix (G{\'o}rski et al. 2005) package.
\end{acknowledgements}

%
\bibliographystyle{aa} 
\bibliography{myReferences} 
%

\begin{appendix} 
\section{Dust polarized data: Smoothing and error estimation}
We use the DR2 Planck HFI 353 GHz Stokes polarization maps
($I_{\rm{S}}$, $Q_{\rm{S}}$ and $U_{\rm{S}}$) obtained from the full
mission with the five full-sky surveys. We smooth the original maps in
order to increase the S/N or, accordingly, to reduce the
variance in pixel.
The price paid is a loss in resolution of the maps altogether with an
increase in beam difference between the diffuse dust emission and the
point-like quasar. We shall thus find the smallest smoothing length that
leads to reliable dust polarization quantities.
We use the smoothing function from the Python HEALPix package
\citep{Gor2005}. This function uses a Gaussian kernel for the
smoothing of input maps and allows us to handle easily polarization
quantities.
We investigated four values of the FWHM
of the Gaussian kernel: 5', 10', 15' and 20'. Owing to the intrinsic map
resolution, this corresponds to an effective resolutions of about 7.03',
11.15', 15.79', and 20.60', respectively.
From these maps we can compute smoothed polarization quantities.

We evaluate the uncertainties in the smoothed maps using a Monte
Carlo approach via the maps of the block-diagonal per-pixel covariance
matrices released by \textit{Planck} at $N_{\rm{side}} = 2048$.
We adopt the gross approximation
that the errors on Stokes parameters are independent of a
Gaussian\footnote{We do not expect a significant effect of this approximation
on our results. The principal argument is that it turned out that we are dominated
and limited by the sampling variance rather than observational uncertainties.}.
We proceed as follows:
(\textit{i})
For each pixel, and for each Stokes parameter $\hat{\mathcal{S}^i}$,
we generate a random realization according to a normal distribution
centered on the data with a standard deviation taken as the square
root of the corresponding diagonal element from the covariance matrix
given in the \textit{Planck} data, $\sqrt{\mathcal{C_{SS}}^i}$.
(\textit{ii})
We proceed to the smoothing of the polarization maps, keeping the
$N_{\rm{side}}$ parameter to 2048.
(\textit{iii})
We extract the Stokes parameters in pixels that contain the quasars.
(\textit{iv}) We run from (\textit{i}) to (\textit{iii})
1000 times. This allows
us to evaluate the mean and the standard deviation for the smoothed
$I_{\rm{S}}$, $Q_{\rm{S}}$ and $U_{\rm{S}}$. As for the errors on the
smoothed quantities, we attribute the values of the corresponding
standard deviations obtained through the simulations. 
Let us note that these errors are likely overestimated since the noise of
the data is accounted twice while generating the simulations. The noise
is indeed already included in the measurements from which we generate
the random realizations and there is no easy way to remove the noise
with confidence.
As a result, the significance of detected correlations quoted in the paper
are lower bounds and, therefore, can be seen as conservative.

From the smoothed $I_{\rm{S}}$, $Q_{\rm{S}}$, and $U_{\rm{S}}$ and
their estimated errors, we compute the smoothed polarized intensities
$P_{\rm{S}}$, the polarization position angles $\psi_{\rm{S}}$, and their
(overestimated) uncertainties ($\sigma_{\rm{P_S}}$, $\sigma_{\rm{\psi_S}}$).

\medskip

For the 355 lines of sight toward the quasars, the fractions that pass
the selection $P_{\rm{S}} \geq \sigma_{\rm{P_S}}$ are 70.1\%, 71.5\%,
80.0\%, and 86.5\% for the FWHM smoothing kernel of 5', 10', 15', and 20'
respectively. As a comparison, only about 60 per cent of these lines of
sight fulfill the above quality criterion if we consider the original maps.
We choose the FWHM value of 15' to conduct our discussion in the core
of the paper as it provides us with a fair compromise between polarized
data quality and relative beam difference between the dust and the quasars.
We do not reject from our analysis the 20 per cent of the data that do
not fulfill the above criterion.
This is because we might be interested in finding those quasars for
which the optical polarization is not too affected by the intervening
Galactic dust. The sight lines corresponding to poorly detected dust
polarized intensity especially represent good candidate regions where
dust has negligible effects on polarization inquiries. Therefore, we
prefer to keep these regions in our analyses. Nevertheless, we investigate the
impact of the inclusion of those lines of sight in the paper and in particular
in Sect.~4.2.
Additionally, and as demonstrated in the next sections, we conducted our
analysis in parallel for the other FWHM values and found consistent
results. Furthermore, in the core of the paper, we took into account the
observational uncertainties while deriving the most important results.

\section{Uniformity of $\Delta_{\rm{S/V}}$ and FWHM values} 
The probabilities given by a one-sample Kolmogorov-Smirnov
test that the distributions of $\Delta_{\rm{S/V}}$ obtained for the FWHM
of 5', 10', 15', and 20' are uniform are found to be $3.95$, $0.04$,
$2.4\,10^{-4}$ and $2.6\,10^{-4}$ per cent.
This illustrates the fact that the departure from uniformity of the
distribution of the relative angles between the quasar optical
polarization vectors and the dust polarization vectors is robust
against the choice of the smoothing value. We checked that this is also
the conclusion reached by means of the other tests used in Sects.~3.2
and~3.3.
The reason why the departure is less significant for small values of the
FWHM is likely because the uncertainties on the dust
polarization angles smear the actual correlation.
Also, the fact that the probabilities in Table~\ref{table:Tab_4} decrease
with increasing FWHM is consistent with the view that the magnetic field
has a high degree of coherence at high Galactic latitudes.

\begin{figure}[h]
        \centering
        \includegraphics[width=\hsize]{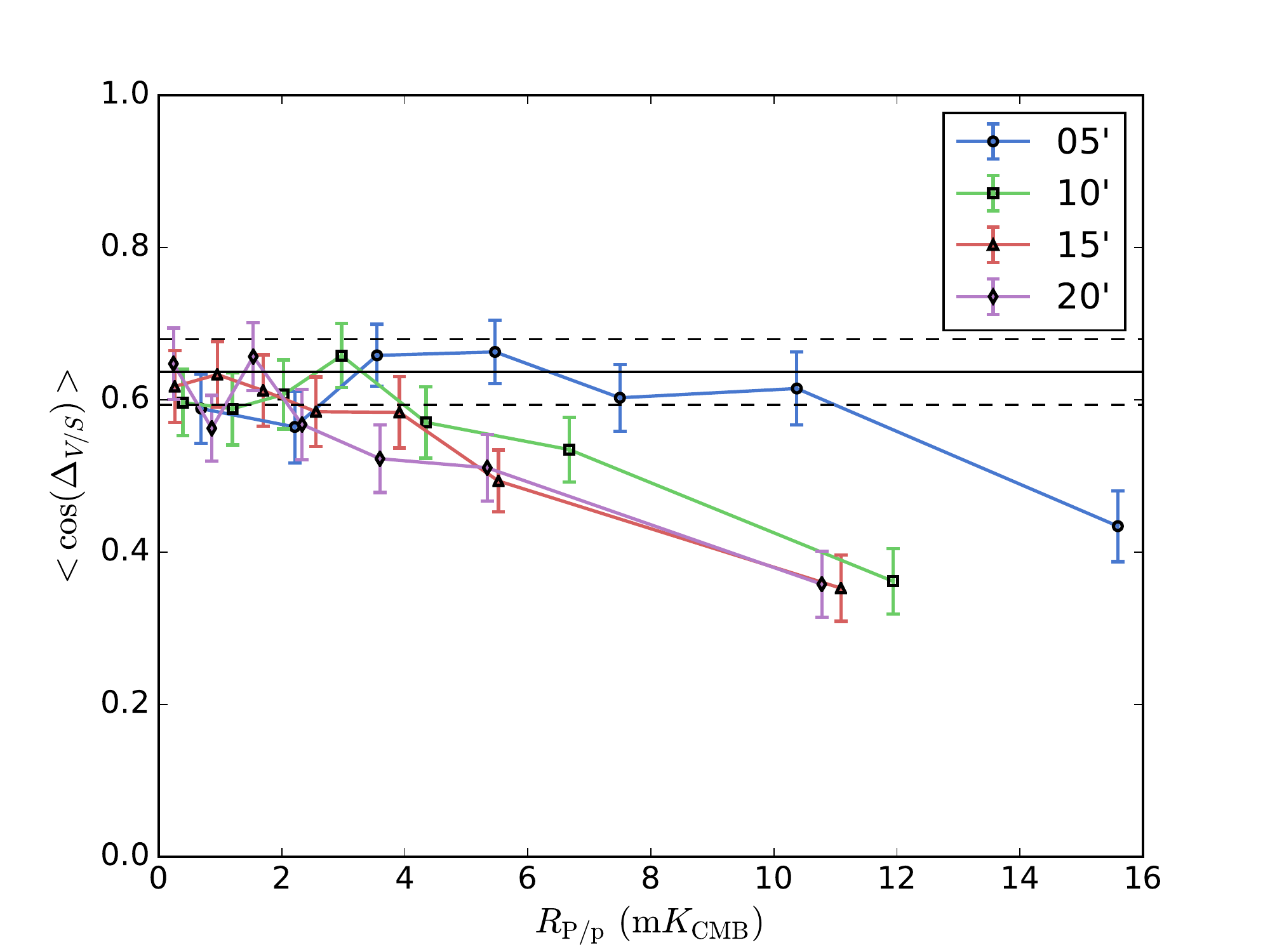}
        \caption{
        Same as for Fig.~\ref{fig:RPpDelta} but for different values
        of the smoothing of the maps.
        Error bars represent the sampling uncertainties in bins
        that we computed as the mean standard errors, i.e., the
        dispersion in bin divided by the square root of the number of data
        points in bin. The significant departure from uniformity in the last
        bin is observed for all tested FWHM values.
        }
        \label{fig:RPpDelta_smAll}
\end{figure}

In Fig.~\ref{fig:RPpDelta_smAll}, we reproduce the results obtained in
Figs.~\ref{fig:RPpDelta} for the four FWHM values.
The trend that we reported in Sect.~3.3 for FWHM = 15' is confirmed.
The larger the values of the polarization ratios, the more likely perpendicular
the quasar optical polarization vectors from the dust polarization vectors.
It is worth mentioning that the reason why the polarization ratios tend to smaller
values when the FWHM increases comes from beam depolarization.
The polarization intensity $P_{\rm{S}}$ can indeed get smaller while
averaging the fluctuating $Q_{S}$ and $U_{S}$ Stokes parameters in the
beam.
The Spearman's rank-order correlation coefficients and the
probabilities that the correlations happen by chance are reported in
Table~\ref{table:Tab_4} for the four FWHM values and two
polarization ratios $R_{\rm{P/p}}$ and $R_{\rm{S/V}}$.

\begin{table}[h]
\caption{Spearman rank-order correlation test between
$\Delta_{\rm{S/V}}$ and the polarization ratios $R_{\rm{P/p}}$
and $R_{\rm{S/V}}$ for the different FWHM values. The values
$\rho$ and $P_{\rm{Spe}}$ are, respectively, the correlation coefficient
computed for the pairs and the two-sided probabilities that such
correlation is due to chance.
}
\label{table:Tab_4}
\centering
\begin{tabular}{l c c c c c}
\hline\hline
\\[-1.5ex]
FWHM            & \multicolumn{2}{c}{$R_{\rm{P/p}}$}    & {}    &\multicolumn{2}{c}{$R_{\rm{S/V}}$} \\
                                & $\rho$        & $P_{\rm{Spe}}$ (\%)           & {}      & $\rho$        & $P_{\rm{Spe}}$ (\%)           \\
\hline
\\[-1.5ex]
5'                              &       0.08            &       $12.44$                         & {}      &               0.11            &       $4.40$                          \\
10'                     &       0.19            &       $0.39$                          & {}      &               0.19            &       $3.5\,10^{-2}$  \\
15'                     &       0.24            &       $5.5\,10^{-4}$  & {}      &               0.23            &       $9.1\,10^{-4}$  \\
20'                     &       0.25            &       $5.5\,10^{-4}$  & {}      &               0.25            &       $2.2\,10^{-4}$  \\[.5ex]
\hline
\end{tabular}
\end{table}

\section{Uniformization of $\Delta_{\rm{S/V}}$ distribution}
\begin{figure}[h]
        \centering
        \includegraphics[width=\hsize]{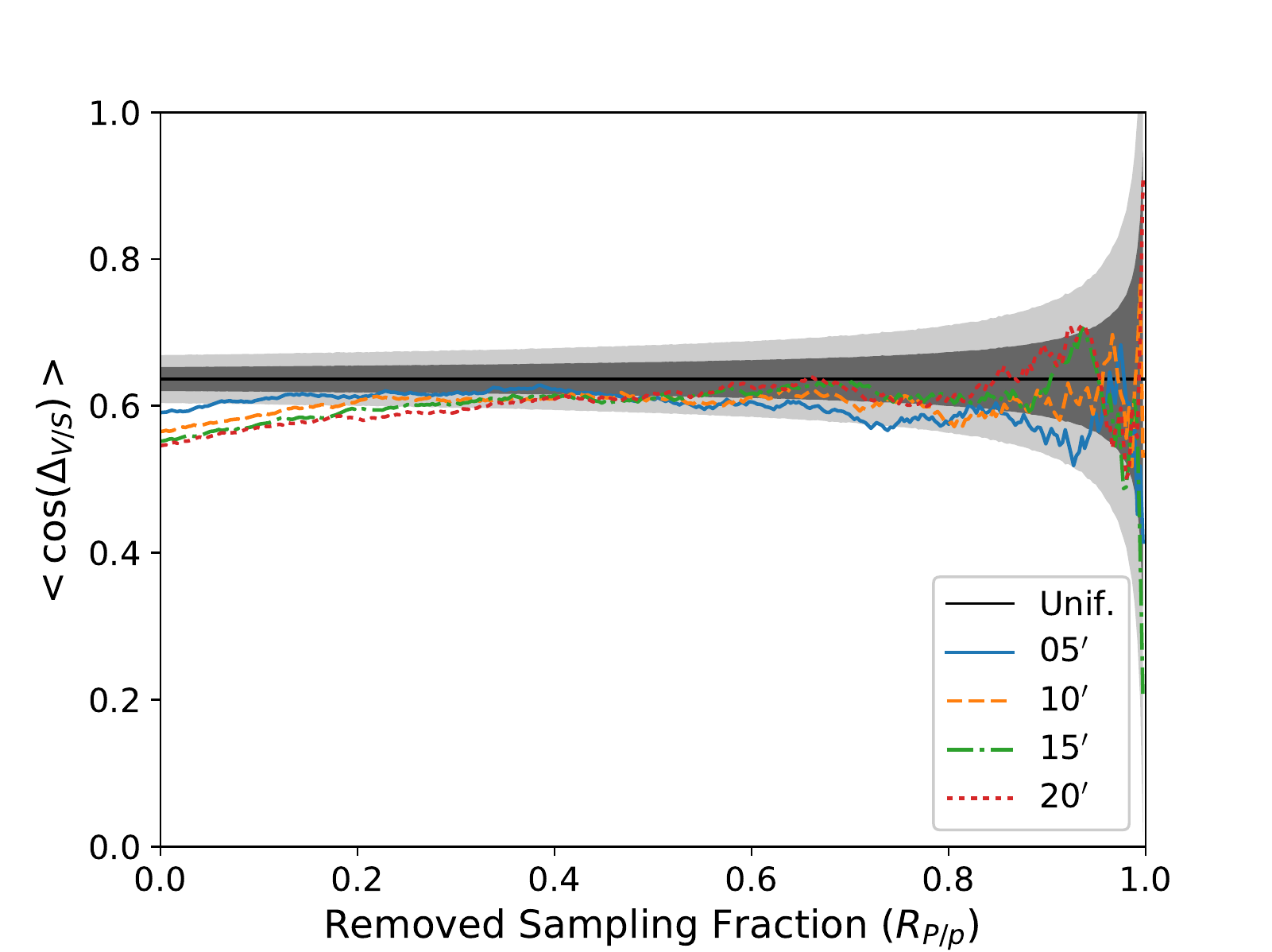}
        \caption{
        Gray shaded regions correspond to the one and two sigma deviations
        around the expected value (black horizontal line), assuming a uniform
        distribution of $\Delta_{\rm{S/V}}$ angles.
        The different lines refer to the different values of the FWHM used to evaluate
        the $R_{\rm{P/p}}$ polarization ratio.
        They represent the $\left\langle \cos(\Delta_{\rm{S/V}} \right\rangle$ values
        from the data as a function of the fraction of the sample that is gradually
        removed while decreasing the $R_{\rm{P/p}}$ selection threshold.
        }
        \label{fig:TruncationRPp_smAll}
\end{figure}

An interesting result obtained in the paper is that the entire quasar optical
polarization sample does not appear to be significantly affected by the Galactic dust.
Removing about 30\% of the sample corresponding to the largest values of
the polarization ratios leads to only marginal deviation of the $\Delta_{\rm{S/V}}$
distribution. This has been set using the
smoothing value of 15'. Similar results are reached with the other
FWHM values. This is illustrated in Fig.~\ref{fig:TruncationRPp_smAll}
where we reproduce Fig.~\ref{fig:SamplingFraction_RPp} (\textit{top}) for the
four tested FWHM values but without considering observational uncertainties.
Taking into account observational uncertainties (not shown on the
figures),
the fraction of the original sample that needs to be truncated to obtain
a $\Delta_{\rm{S/V}}$ distribution that agrees with uniformity at the $2\sigma$
level is between about 20 and 40 per cent for all the studied FWHM values.
That is, 80 to 60 per cent of the quasar optical polarization sample appear
sufficiently decoupled from the Galactic dust polarization such that we
can trust the optical polarization to be intrinsic to the quasars and not too
contaminated by intervening Galactic dust.

\section{ Polarization ratio $R_{\rm{S/V}}$}
In Sect.~3.1, we discussed two polarization ratios $R_{\rm{P/p}}$ and
$R_{\rm{S/V}}$ that mix observable quantities from the optical and
submillimeter. As defined, $R_{\rm{P/p}}$ is only sensitive to the
polarizing dust grains. The $R_{\rm{S/V}}$ ratio, however, also encodes
the effect of nonaligned grains on optical background source light.

\begin{figure}[h]
        \centering
        \includegraphics[width=\hsize]{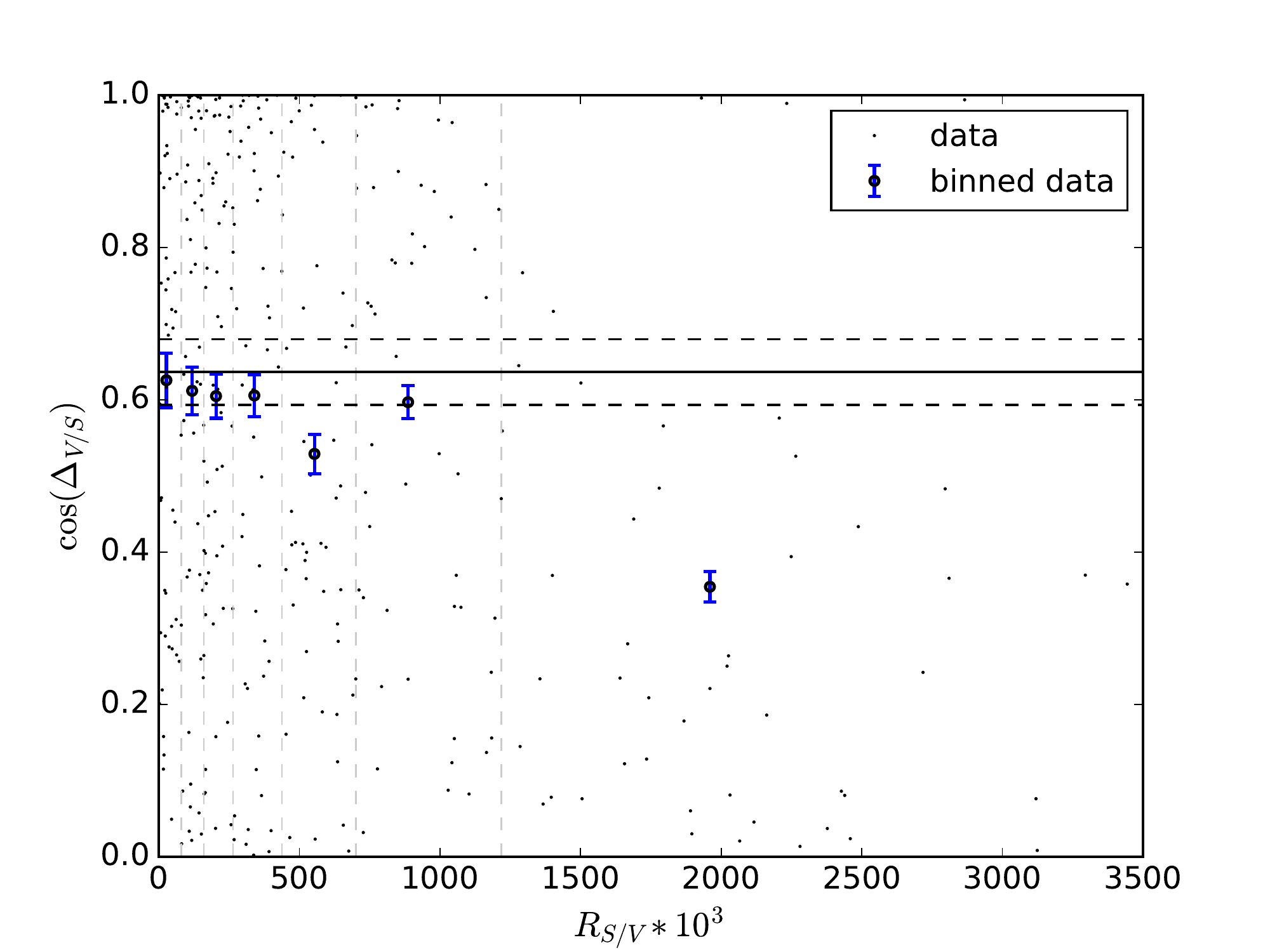}
        \caption{
        Same as for Fig.~\ref{fig:RPpDelta} but for the pairs
        $(\cos(\Delta_{\rm{S/V}}),\, R_{\rm{S/V}})$.
        }
        \label{fig:RSVDelta}
\end{figure}

\begin{figure}[h]
        \centering
        \includegraphics[width=\hsize]{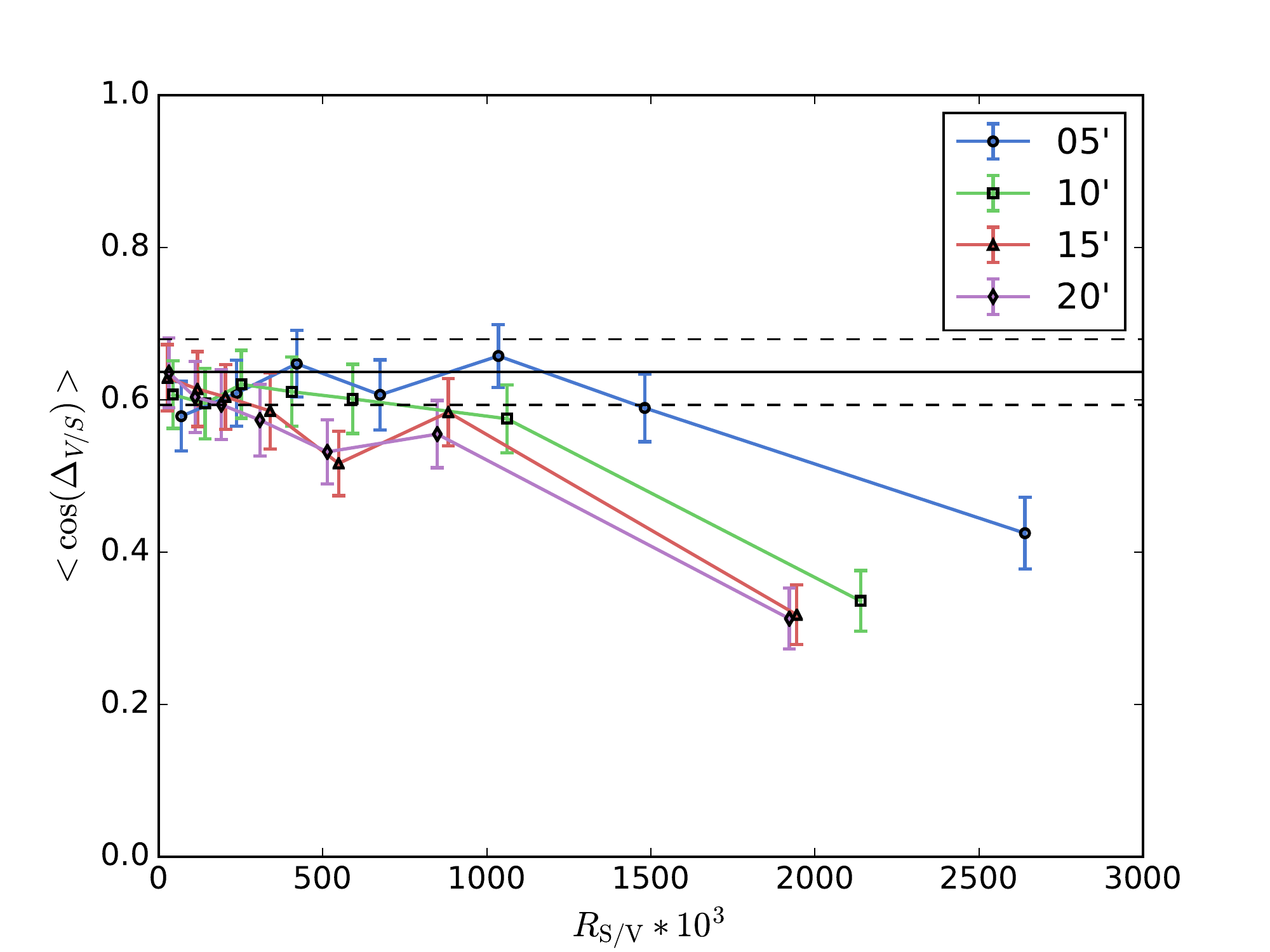}
        \caption{
        Same as for Fig.~\ref{fig:RPpDelta_smAll} but for the pairs
        $(\cos(\Delta_{\rm{S/V}}),\,R_{\rm{S/V}})$.
        }
        \label{fig:RSVDelta_smAll}
\end{figure}

\begin{figure}[h]
        \centering
        \includegraphics[width=\hsize]{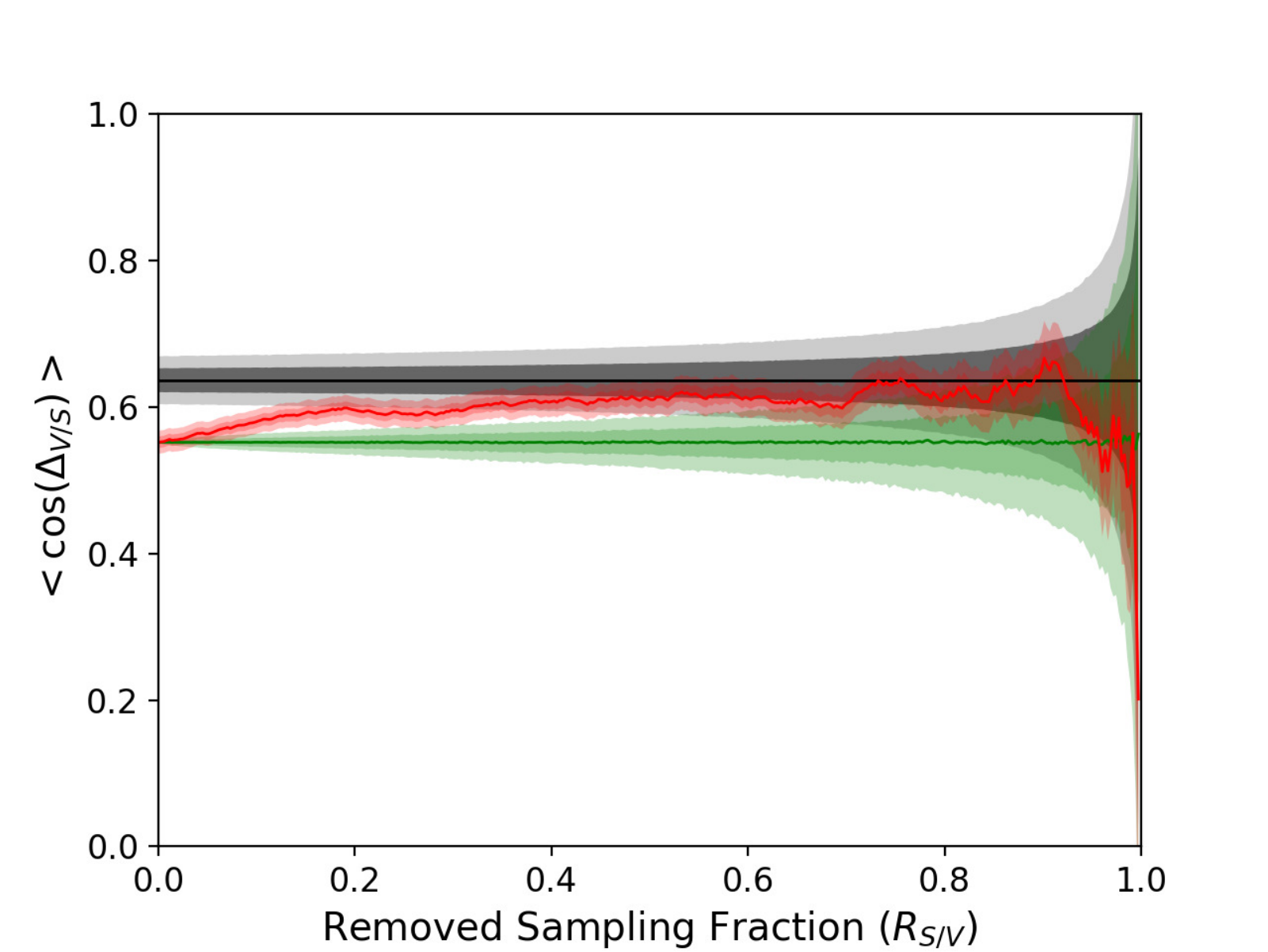}
        \caption{
        Same as for Figs.~\ref{fig:SamplingFraction_RPp} and ~\ref{fig:TruncationRPp}
        but with the polarization ratio $R_{\rm{S/V}}$.
        The mean and the 1$\sigma$ and 2$\sigma$ regions of the distribution of 
        $\left\langle \cos(\Delta_{\rm{S/V}} \right\rangle$ obtained with random
        truncation is illustrated in green.
        }
        \label{fig:SamplingFraction_RSV}
\end{figure}

\begin{figure}
        \centering
        \includegraphics[width=\hsize]{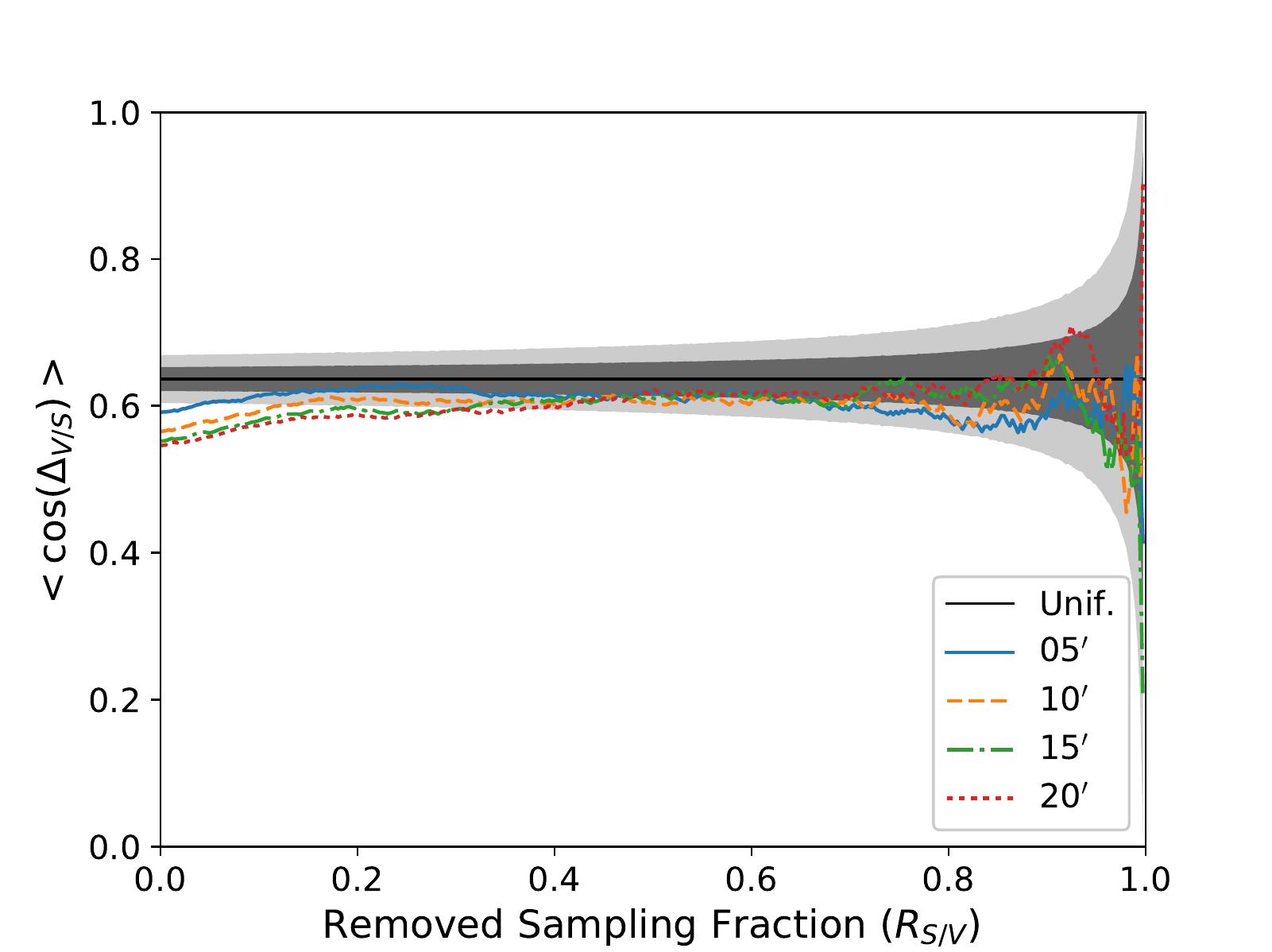}
        \caption{
        Same as for Fig.~\ref{fig:TruncationRPp_smAll} but for the
        $R_{\rm{S/V}}$ polarization ratio.
        }
        \label{fig:TruncationRSV_smAll}
\end{figure}

For clarity, we presented in the core of the paper only the results of the
analyses obtained with the $R_{\rm{P/p}}$ ratio.
As we performed in parallel all the analyses considering $R_{\rm{S/V}}$,
we show the main results in Figs.~\ref{fig:RSVDelta},
\ref{fig:RSVDelta_smAll},
\ref{fig:SamplingFraction_RSV}, and~\ref{fig:TruncationRSV_smAll}.
These correspond to Figs.~\ref{fig:RPpDelta},
\ref{fig:RPpDelta_smAll}, \ref{fig:SamplingFraction_RPp},
and~\ref{fig:TruncationRPp_smAll}, respectively.
As illustrated by these figures, we obtained very consistent results
considering either $R_{\rm{S/V}}$ or $R_{\rm{P/p}}$. This could have
been expected in light of the observed correlation between the  two
polarization ratios in Fig.~\ref{fig:RPp-RSV}.

\end{appendix}

\end{document}